\pdfoutput=1

\documentclass[journal,onecolumn]{IEEEtran}

\usepackage{subcaption}

\newdimen\nodeDist
\usepackage{pgfplots}
\pgfplotsset{width=7cm, compat=1.10}
\usepgfplotslibrary{fillbetween}
\pgfmathdeclarefunction{poly}{0}{\pgfmathparse{-x^3+5*(x^2)-3*x-3}}

\usepackage{tcolorbox}

\usepackage{rotating}
\usepackage{amsfonts}
\usepackage{amsmath}
\usepackage{calc}

\usepackage{enumitem}

\usepackage{color}
\usepackage{xcolor}

\usepackage{colortbl}

\usepackage{booktabs}
\usepackage{placeins}

\usepackage{mdframed}

\definecolor{sigLarge}{gray}{0.6}
\definecolor{sigMedium}{gray}{0.75}
\definecolor{sigSmall}{gray}{0.9}

\usepackage{tikz}
\usetikzlibrary{positioning, backgrounds, shapes.geometric} 
\tikzstyle{vertexStrong} = [draw, fill=sigLarge, text=white, regular polygon, regular polygon sides=4]
\tikzstyle{vertexMedium} = [draw, fill=sigMedium, text=white, regular polygon, regular polygon sides=4]
\tikzstyle{vertexSmall} = [draw, fill=sigSmall, text=white, regular polygon, regular polygon sides=4]

\newcommand{\vertexStrong}{\protect\tikz{\protect\node[vertexStrong, scale=0.7](0,0) {};}}
\newcommand{\vertexMedium}{\protect\tikz{\protect\node[vertexMedium, scale=0.7](0,0) {};}}
\newcommand{\vertexSmall}{\protect\tikz{\protect\node[vertexSmall, scale=0.7](0,0) {};}}

\usepackage[ruled]{algorithm2e} 

\SetAlFnt{\small}
\SetAlCapFnt{\small}
\SetAlCapNameFnt{\small}
\SetAlCapHSkip{0pt}
\IncMargin{-\parindent}

\begin{document}

\newcommand{\todo}[1]{\textcolor{red}{TODO #1}}
\newcommand{\new}[1]{\textcolor{blue}{#1}}

\newcommand{\changeReview}[1]{\textcolor{cyan}{#1}}
\newcommand{\delete}[1]{\textcolor{gray}{#1}}

\tcbset{before upper=\hfill\scshape, before lower=\itshape, sharp corners=all, lower separated=false, colframe=gray!90!black, colback=gray!25, size=small,leftrule=0.2mm, rightrule=0.2mm,toprule=0.2mm,bottomrule=0.2mm}

\newcommand{\Recommendation}[1]{
\begin{tcolorbox}
 Recommendation
 \tcblower
 #1
\end{tcolorbox}
}

\mdfdefinestyle{RQFrame}{%
    linecolor=white,
    outerlinewidth=0pt,
    innertopmargin=8pt,
    innerbottommargin=0pt,
    innerrightmargin=0pt,
    innerleftmargin=-1pt,
        leftmargin = 0pt,
        rightmargin = 0pt
        }

%
\title{Predicting Performance of Software Configurations:\\ There is no Silver Bullet}

\author{\IEEEauthorblockN{Alexander Grebhahn}
\IEEEauthorblockA{
University of Passau\\
Passau, Germany\\
}
\and
\IEEEauthorblockN{Norbert Siegmund}
\IEEEauthorblockA{Bauhaus-University Weimar\\
Weimar, Germany\\
}
\and
\IEEEauthorblockN{Sven Apel}
\IEEEauthorblockA{
Saarland University\\
Saarbr\"ucken, Germany\\
}

}

\IEEEtitleabstractindextext{%
\begin{abstract}

Many software systems offer configuration options to tailor their functionality and non-functional properties (e.g., performance). 
Often, users are interested in the (performance-)optimal configuration, but struggle to find it, due to missing information on influences of individual configuration options and their interactions. 
In the past, various supervised machine-learning techniques have been used to predict the performance of all configurations and to identify the optimal one.

In the literature, there is a large number of machine-learning techniques and sampling strategies to select from. 
It is unclear, though, to what extent they affect prediction accuracy.
We have conducted a comparative study regarding the mean prediction accuracy when predicting the performance of all configurations considering 6 machine-learning techniques, 18 sampling strategies, and 6 subject software systems.

We found that both the learning technique and the sampling strategy have a strong influence on prediction accuracy.
We further observed that some learning techniques (e.g., random forests) outperform other learning techniques (e.g., k-nearest neighbor) in most cases.
Moreover, as the prediction accuracy strongly depends on the subject system, there is no combination of a learning technique and sampling strategy that is optimal in all cases, considering the tradeoff between accuracy and measurement overhead, which is in line with the famous no-free-lunch theorem.

\end{abstract}
}

%


%
%


\maketitle

\IEEEdisplaynontitleabstractindextext

\section{Introduction}
Software systems often provide numerous configuration options and tuning possibilities to adjust them to specific hardware platforms, operating systems, and use cases.
Besides adjusting functionality, configuration options can also have a significant influence on non-functional properties such as the performance of the system (e.g., response time)~\cite{SGA+15}.
In fact, in many domains, such as in server software and high-performance computing, it is essential to tune a system not solely based on the required functionality, but also regarding performance or other non-functional properties.
However, identifying the best performing configuration for a specific platform and use case and, tuning the system accordingly, is a non-trivial task~\cite{Xu:2015:HYG:2786805.2786852}.
To this end, various performance modeling techniques have been used, which can be classified in analytical and empirical techniques. 
Analytical techniques use domain knowledge about the internal functionality of the system, while empirical performance modeling techniques rely on measurements. 
Although analytical performance models were created for different systems (see for example, \cite{DBLP:conf/ics/GahvariBSYJG11}), they heavily rely on abstraction, which might have an influence on their accuracy in practical settings~\cite{books/daglib/0076234}.
Empirical performance models require a set of configurations being measured beforehand, which can become costly and time-consuming.
Since a direct measurement of all configurations of a software system is rarely feasible in practice~\cite{Guo2013}, we need a mechanism that accurately predicts their performance, after measuring only a small number of them.

Under the umbrella of supervised machine learning, there are many techniques that can be used for empirical performance modeling and performance prediction.
Prominent examples are \textit{classification and regression trees}~\cite{cart84}, \textit{support-vector regression}~\cite{basak2007support}, and \textit{multiple linear regression}~\cite{anderson2003introduction}.
Each of these techniques has its pros and cons, depending on the non-functional behavior of the software system, the available resources in terms of available measurements, the intended use case (i.e., prediction vs. understanding), and so on~\cite{KSK+18}.
The immediate question that arises here is, which technique shall be used when?
There are already some comparisons available~\cite{Valov:2015:ECR:2791060.2791069, Garcia-Gutierrez2014}, which demonstrate the continuous improvement of this field over the last years; novel techniques are usually more accurate and require fewer measurements. 
However, there is one important and systematic factor ignored in all of these studies: \emph{the influence of the distribution of the learning set used for learning}.

Prediction accuracy---that is, the ratio between the measured performance value and the predicted performance value of a configuration---strongly depends on the configurations of the software system that are selected for learning~\cite{SGA+15}.
For example, if a configuration option with a relevant influence is not represented in the learning set, then the technique will not be able to accurately predict configurations with that option enabled.
The same holds for interactions among configuration options, that is, specific performance behaviors that occur only when a specific set of options in enabled in combination.
Thus, the learning set (a.k.a. sample set) is equally important for prediction accuracy as the learning technique~\cite{KaGrSi+19}.


In the recent years, many machine-learning techniques have been used in combination with specific sampling strategies to predict performance of configurations of software systems.
For example, Guo et al. used CART in combination with random sampling to learn performance models~\cite{Guo2013}. 
In follow-up work, they combined CART with feature frequency sampling to improve accuracy~\cite{SGS+15}. 
Siegmund et al. used coverage-oriented sampling strategies, such as the Plackett Burmann Design~\cite{PlackettBurman1946} and T-Wise sampling, in combination with multiple linear regression to learn performance models~\cite{SGA+15}.
In another line of research, Nair et al. used spectral learning in combination with actively sampled learning sets~\cite{DBLP:journals/corr/abs-1801-02175, DBLP:journals/corr/NairMSA17}.
Kaltenecker et al.~\cite{KaGrSi+19} proposed a new sampling strategy covering the whole configuration space as equally as possible and demonstrated that using the newly proposed strategy can significantly improve prediction accuracy of multiple linear regression.

All of this work demonstrates the applicability and usefulness of certain combinations of machine learning and sampling in specific settings.
However, it remains unclear whether using another learning technique might lead to even more accurate (or worse) predictions or whether the used sampling strategy (not the learning technique) was responsible for the good results.
In general, mostly random samples have been used in experiments. 
So, it is actually unclear whether selecting more or less configurations, or configurations following a specific probability distribution, for learning would influence the accuracy of prediction.

Summarizing the state of the art, there is no systematic and comprehensive understanding of which sampling strategy is appropriate for which learning technique and vice versa.
To shed light on this issue, we have conducted an extensive empirical study.
In particular, we compare a set of 18 different sampling strategies and 6 supervised machine-learning techniques regarding their prediction accuracy on 6 real-world subject software systems; we compare all combinations regarding predicting the performance of \textit{all} configurations of subject systems. 
Overall, this resulted in more than 15\,000 experiments.
To make this comparison fair, we perform hyper-parameter optimization for each of the experiments to determine for each experiment the optimal setting\footnote{We performed one parameter optimization on each combination of a machine-learning technique with a binary and a numeric sampling strategy on each subject system.}.
The study compares substantially more subject systems, learning techniques, and sampling strategies than any other study so far.


Overall, we make the following contributions: 
\begin{itemize}
  \item We provide a structured comparison of the influence of different combinations of machine-learning techniques and sampling strategies on the accuracy of performance prediction of software configurations.
  \item We show that the selection of the optimal combination of learning technique and sampling strategy with respect to prediction accuracy and measurement overhead depends on the subject system.
  Although there is no combination leading to the most accurate predictions for all subject systems, we see significant differences between different machine learning techniques or sampling strategies.
\end{itemize}

Because of space limitations, we present only the aggregated results over all subject systems in this work and provide more detailed analyses at our supplementary Web site\footnote{\label{webPage}https://www.se.cs.uni-saarland.de/projects/splconqueror/expDesign/expDesign.html}.

\section{Foundations, Problem Statement \& Research Questions}

In this section, we define basic terminology and present our research questions.

\subsection{Foundations}


\textit{Configurable Systems:}
A configurable software system $s$ provides a set $O_s$ of configuration options.
We differentiate between binary options $O_{\mathit{Bin}}$ and numeric options $O_{\mathit{Num}}$, that is, $O_s = O_{\mathit{Bin}} \cup O_{\mathit{Num}}$.
While binary options can take only two values, 0 and 1 ($\mathit{dom} : O_{\mathit{Bin}}\rightarrow \{0,1 \}$), numeric options are defined by a numeric value domain ($\mathit{dom} : O_{\mathit{Num}}\rightarrow \mathbb{N}$).
For example, the decision of whether using a specific index structure in a database system is represented by a binary option, whereas the choice of the page size of the database is specified by a numeric option.
We further define \textit{Constr} as the set of constraints, specifying valid value combinations of configuration options.
For example, the sum of the two numeric configuration options \textit{pre} and \textit{post} of the \textit{HSMGP} system, which we consider in this work, has to be larger than~0, although both their value domains include~0.
A \textit{valid assignment} of every configuration option to a value from its respective domain yields a valid configuration $c \in C(s)$. 


\textit{Performance prediction:}
In our study, we use \textit{regression} to predict a real valued output (i.e., performance) for each configuration of a given configurable system.
With regression, we learn a predictor $\Pi : C \rightarrow \mathcal{R}$ that takes a configuration $c \in C(s)$ as input and predicts the corresponding performance $p \in \mathcal{R}$.
To learn the predictor $\Pi$, we rely on a leaning set $\mathcal{L} \subset C(s)$ of configurations.
Once learned, the predictor can be used to predict the performance of all configurations of the considered subject system.
For example, after measuring only 250 configurations of the subject software system \textit{VP9}, we can predict all 216\,000 configurations provided by this system.

\subsection{Problem Statement \& Research Questions}

There is a multitude of different supervised machine-learning techniques that can be used in our setting (to learn $\Pi$).
In previous work, it has been shown that individual techniques can predict performance with a high accuracy~\cite{SGA+15,Guo2013, DBLP:journals/corr/NairMSA17}, but it is unclear whether some techniques outperform other techniques, in general, or whether this depends on the subject system.
Thus, preferring one technique over the other might lead to a loss in prediction accuracy.

All supervised machine-learning techniques rely on a learning set of configurations.
Often, a simple random selection of configurations is performed to create this learning set~\cite{Guo2013, temple:hal-01467299}.
In addition, there is a large number of different structured sampling strategies that can be used to select the learning set, for example, a learning set that considers all options equally often~\cite{myers2009response}.
More generally, a sampling strategy selects configurations based on specific criterion (e.g., one configuration for each option, where only the option of interest is selected and all other options are deselected, if possible) to enable deriving specific properties of the subject system, such as individual influences or interactions among a specific number or kind of configuration options.
However, again, it is unclear which of these structured or random sampling strategies should be used and in combination with which learning technique.
Beside merely predicting performance, it is important to achieve stable results, when for example, different learning sets or machine-learning techniques on the same learning set are used.
This is because stable results indicate that the learning technique or sampling strategy can also lead to similar accuracies on other subject systems.

Our goal is to study the influence of individual combinations of machine-learning techniques and sampling strategies on the accuracy when predicting the performance of all configurations of a configurable software system. 
To this end, we formulate separate research questions to first consider individual aspects of the machine-learning techniques and sampling strategies before addressing their combined effects on prediction accuracy.

\paragraph{Machine-learning techniques}

Different learning techniques have been applied to performance prediction of configurable software systems.
But, it is largely unknown which is superior and when:
\begin{mdframed}[style=RQFrame,nobreak=true]
\textbf{RQ1.1:} Are there specific learning techniques that are superior to all other techniques?
\end{mdframed}
\begin{mdframed}[style=RQFrame,nobreak=true]
\textbf{RQ1.2:} Which learning techniques yield the most stable prediction accuracies with respect to varying learning sets?
\end{mdframed}

\paragraph{Sampling strategies}

The accuracy of a learning technique depends on the learning set, and there are different strategies to produce such a set.
We are interested in which sampling strategy leads to the most accurate and stable predictor, independently of the learning technique:
\begin{mdframed}[style=RQFrame,nobreak=true]
\textbf{RQ2.1:} Which sampling strategy leads to the most accurate predictions?
\end{mdframed}
\begin{mdframed}[style=RQFrame,nobreak=true]
\textbf{RQ2.2:} Which sampling strategy leads to most stable prediction accuracies?
\end{mdframed}

\paragraph{Machine-learning techniques \& sampling strategies}

So far, we looked at learning techniques and sampling strategies individually.
The following research questions are concerned with whether \textit{specific combinations} of machine-learning techniques and sampling strategies are superior:
\begin{mdframed}[style=RQFrame,nobreak=true]
\textbf{RQ3.1:} Which combination of machine-learning technique and sampling strategy produces the most accurate results?
\end{mdframed}
\begin{mdframed}[style=RQFrame,nobreak=true]
\textbf{RQ3.2:} Are there specific aspects of sampling strategies, such as interaction coverage, that lead to more accurate performance predictions of learning techniques that take advantage of them?
\end{mdframed}

\section{Machine-Learning Techniques}\label{sec:machineLearning}

Predicting a real-valued output variable based on a number of input variables is a well known problem in machine learning, which has led to a number of techniques.
These techniques rely on different strategies for learning (e.g., some perform an input transformation while others use trees to partition the configuration space recursively).
Despite considerable research~\cite{Guo2013, SGA+15, KSK+18}, it is not clear to what extent the different strategies are applicable for performance prediction of configurable software systems, as these systems provide distinct patterns of interactions~\cite{SSA+17}.

In our study, we consider 6 prominent supervised machine-learning techniques (Multiple Linear Regression, Classification and Regression Trees, Random Forest, Support Vector Machines, Kernel Ridge Regression, and k-Nearest-Neighbors Regression), which we will explain shortly.

While not being comprehensive, we have selected the most prominent and reasonable ones, based on current practice. 
Methods, such as deep neural networks can be used, too, but it has been shown that well optimized standard techniques often outperform those hyped techniques~\cite{DBLP:conf/sigsoft/FuM17}. 
More importantly, the chosen techniques are known that they work on small learning sets, which is a crucial precondition for our settings, see for example~\cite{SGA+15,Guo2013}.

Each technique usually comes with its own set of configuration options, called \emph{hyper-parameters}.
Hyper-parameters are essential for achieving an accurate prediction and, therefore, we optimize in our study also these parameters (see Section~\ref{sec:mlParamOpt}).
In Table~\ref{tab:mlOptimization}, show the machine-learning techniques including their parameters we consider in this work.

\subsection{Multiple Linear Regression \& Feature Subset Selection}

First, we consider an approach that uses \textit{multiple linear regression} (MR) to derive a mathematical function that describes all relevant influences of the configuration options and their interactions on performance~\cite{SGA+15}.
The following function gives an example with one binary option $o_{\mathit{Bin}_1}$ and one numeric option $o_{\mathit{Num}_1}$:

\begin{equation}
\begin{array}{l}

\Pi(c) = \overbrace{c_1}^{\pi_1} + \overbrace{c_2 \cdot o_{\mathit{Bin}_1}}^{\pi_2} + \overbrace{c_3 \cdot o_{\mathit{Bin}_1} \cdot o_{\mathit{Num}_1} \cdot o_{\mathit{Num}_1} }^{\pi_3} + ...

\end{array}
\end{equation}

The predictor function consists of multiple, linearly combined terms ($\pi_1,...,\pi_3$), in which each term expresses the influence of either a single configuration option (e.g., $\pi_2$) or a combination of options (an interaction, e.g., $\pi_3$) on performance.
The strength of the influence is specified by the constant $c_i$.
Note that, for numeric options, there can also be non-linear influences, such as a quadratic influence, as in term $\pi_3$.

In the worst case, the number of interactions is exponential in the number of configuration options.
For numeric configuration options, the number of possible interactions is even higher, due to possible non-linear influences.
This high dimensionality of possible terms in the predictor function makes it necessary to select only the most relevant ones, which can be realized with feature subset selection.

With \textit{feature subset selection}, the predictor function is iteratively constructed by adding an additional term each iteration such that the prediction error for a validation set is reduced most.
This process is repeated until a certain stopping criteria is reached.
For more details on this approach, we refer to Siegmund et al.~\cite{SGA+15}.

\subsection{Trees and Forests}

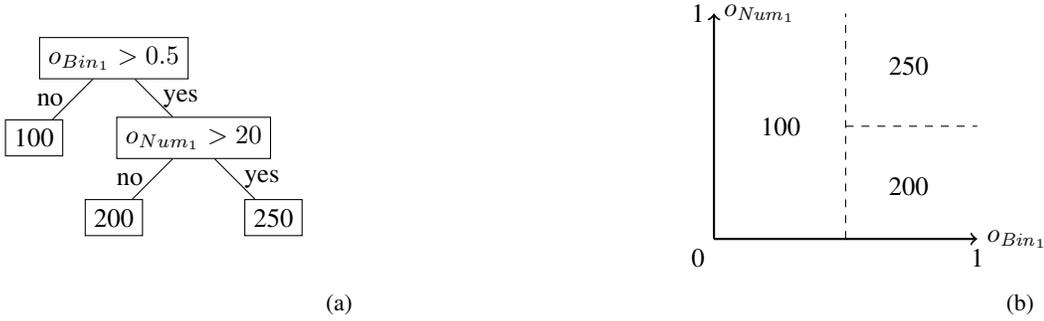
\begin{figure}
\begin{subfigure}[b]{.49\linewidth}
\begin{tikzpicture}[node/.style={draw, rectangle}]
    \node [node] (A) {$o_{Bin_1} > 0.5$};
    \path (A) ++(-135:\nodeDist) node [node] (B) {100};
    \path (A) ++(-45:\nodeDist) node [node] (C) {$o_{Num_1} > 20$};
    \path (C) ++(-135:\nodeDist) node [node] (D) {200};
    \path (C) ++(-45:\nodeDist) node [node] (E) {250};

    \draw (A) -- (B) node [left,pos=0.5] {no}(A);
    \draw (A) -- (C) node [right,pos=0.5] {yes}(A);
    \draw (C) -- (D) node [left,pos=0.5] {no}(A);
    \draw (C) -- (E) node [right,pos=0.5] {yes}(A);
\end{tikzpicture}
\vspace{1.5em}
\caption{}
\end{subfigure}
\begin{subfigure}[b]{.495\linewidth}
\begin{tikzpicture}
    \draw[->,thick] (0,0)--(3.5,0) node[right]{$o_{\mathit{Bin}_1}$};
    \draw[->,thick] (0,0)--(0,3) node[right]{$o_{\mathit{Num}_1}$};
    
    \draw[dashed] (1.75,0)--(1.75,3) node[right]{};
    \draw[dashed] (1.75,1.5)--(3.5,1.5) node[right]{};
    
    \draw (0.5,1.5) node[right]{100};
    \draw (2.2,0.7) node[right]{200};
    \draw (2.2,2.3) node[right]{250};
    
    \draw (0,-0.25) node[below,left]{0};
    \draw (3.5,0) node[below]{1};
    \draw (0,3) node[left]{1};
\end{tikzpicture}
\caption{}
\end{subfigure}

\caption{(a) A simple regression tree based on two options $o_{\mathit{Bin}_1}$ and $o_{\mathit{Num}_1}$ (b) and its corresponding recursive partition of the configuration space.}
\label{fig:cart}
\end{figure}

\textit{Classification and Regression Trees (CART)} perform a recursive partitioning on the configuration space in terms of decisions on the value of configuration options~\cite{cart84}.
This partitioning can be represented in a binary tree.
In Figure~\ref{fig:cart}, we give an example of a partition and the respective regression tree, considering the influence of two configuration options $o_{\mathit{Bin}_1}$ and $o_{\mathit{Num}_1}$.
The partitioning of the configuration space leads to three disjoint regions being represented as leaf nodes in the tree.
To predict a configuration's performance value, the tree is traversed according to the values defined in the configuration until a leaf node is reached, which stores the corresponding performance value. 
For example, a configuration with $o_{\mathit{Bin}_1} = 1$ and $o_{\mathit{Num}_1} = 30$ has a predicted performance value of 250.

There are different techniques to select the configuration option being used in a partition of the configuration space~\cite{loh2011classification}.
The most common technique selects the option that minimizes the entropy of the performance values.
That is, the most relevant configuration options are selected first.

It is also possible to learn multiple regression trees in parallel, which in combination gives rise to a \textit{Random Forest (RF)}.
In a random forest, each subtree has been learned on a subset of the learning set or on a subset of configuration options. 
This technique is more robust against overfitting and usually generalizes better to new configurations, because the generalization error of one tree is compensated by other trees of the forest~\cite{Breiman2001}.
To predict one configuration using the forest, the average value of the predictions of all trees is used.


\subsection{Support Vector Regression} \label{sec:svM}

\textit{Support vector machines} apply a transformation on the configuration space, using a kernel function, to be able to use linear regression in the transformed high-dimensional space~\cite{vapnik1995nature}.
This transformation needs to be chosen a priori, but identifying a suitable transformation is not trivial.
In our study, we focus on two different kinds of support vector regression.

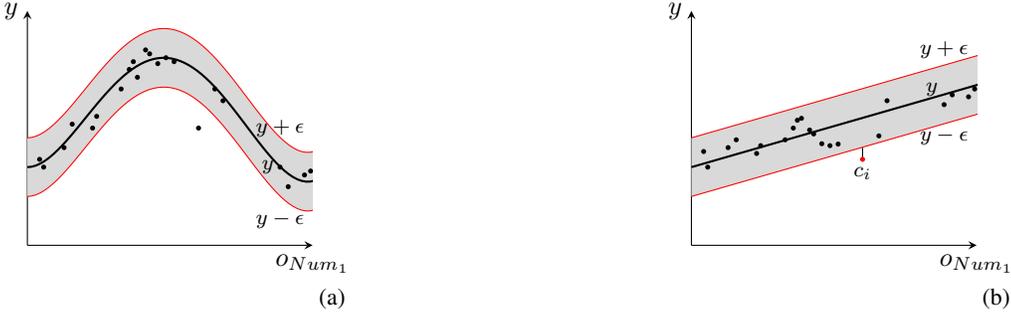
\begin{figure}
\vspace{5mm}
\begin{subfigure}[b]{.48\linewidth}
\begin{tikzpicture}[scale=0.7]
  \begin{axis}[
    xlabel={$o_{Num_1}$},
    ylabel={$y$},
    ylabel style={rotate=-90},
    xticklabel=\empty,
    yticklabel=\empty,
    ticks=none,
    label style ={at={(ticklabel cs:1)}},
    axis y line = left,
    axis x line = bottom,
    samples     = 160,
    domain      = 0:17,
    xmin = 0, xmax = 7,
    ymin = 0, ymax = 6,
  ]

  \addplot[name path=upper, red, mark=none] {(x * x) - (0.3 * x * x * x) + (0.023 * x * x * x * x) - (0.0001 * x* x * x * x * x) + 2.75};
  \node at (62,3) {\footnotesize $y + \epsilon$};
  \addplot[name path=lower, black,thick, mark=none] {(x * x) - (0.3 * x * x * x) + (0.023 * x * x * x * x) - (0.0001 * x* x * x * x * x) + 2};
  \node at (59,2) {\footnotesize $y$};
  \addplot[name path=lower, red, mark=none] {(x * x) - (0.3 * x * x * x) + (0.023 * x * x * x * x) - (0.0001 * x* x * x * x * x) + 1.25};
  \node at (62,0.65) {\footnotesize $y - \epsilon$};
  
  \addplot[fill=gray!30] fill between[ 
    of = upper and lower
  ];
    
  \node [circle, scale=0.2, fill=black] at (3,2.2) {};
  \node [circle, scale=0.2, fill=black] at (4,2) {};
  \node [circle, scale=0.2, fill=black] at (9,2.5) {};
  \node [circle, scale=0.2, fill=black] at (11,3.1) {};
  \node [circle, scale=0.2, fill=black] at (16,3.0) {};
  \node [circle, scale=0.2, fill=black] at (17,3.3) {};
  \node [circle, scale=0.2, fill=black] at (23,4) {};
  \node [circle, scale=0.2, fill=black] at (25,4.5) {};
  \node [circle, scale=0.2, fill=black] at (26,4.7) {};
  \node [circle, scale=0.2, fill=black] at (27,4.3) {};
  \node [circle, scale=0.2, fill=black] at (29,5) {};
  \node [circle, scale=0.2, fill=black] at (30,4.9) {};
  \node [circle, scale=0.2, fill=black] at (32,4.65) {};
  \node [circle, scale=0.2, fill=black] at (34,4.8) {};
  \node [circle, scale=0.2, fill=black] at (36,4.7) {};
  \node [circle, scale=0.2, fill=black] at (46,4.0) {};
  \node [circle, scale=0.2, fill=black] at (48,3.7) {};
  \node [circle, scale=0.2, fill=black] at (62,2.0) {};
  \node [circle, scale=0.2, fill=black] at (64,1.5) {};
  \node [circle, scale=0.2, fill=black] at (69.5,1.9) {};
  \node [circle, scale=0.2, fill=black] at (68,1.8) {};
                        
  \node [circle, scale=0.2, fill=black] at (42,3.0) {};
  
\end{axis}
\end{tikzpicture}
\caption{}
\end{subfigure}
\begin{subfigure}[b]{.48\linewidth}
\begin{tikzpicture}[scale=0.7]
  \begin{axis}[
    xlabel={$o_{Num_1}$},
    ylabel={$y$},
    ylabel style={rotate=-90},
    xticklabel=\empty,
    yticklabel=\empty,
    ticks=none,
    label style ={at={(ticklabel cs:1)}},
    axis y line = left,
    axis x line = bottom,
    samples     = 160,
    domain      = 0:17,
    xmin = 0, xmax = 7,
    ymin = 0, ymax = 6,
  ]

  \draw (42,220) -- (42,250);

  \addplot[name path=upper, red, mark=none] { 0.3 * (x)  + 2.75};
  \node at (62,500) {\footnotesize $y + \epsilon$};
  \addplot[name path=lower, black,thick, mark=none] {0.3 * (x ) + 2};
  \node at (59,400) {\footnotesize $y$};
  \addplot[name path=lower, red, mark=none] {0.3 * (x ) + 1.25};
  \node at (62,280) {\footnotesize $y - \epsilon$};
  
  \addplot[fill=gray!30] fill between[ 
    of = upper and lower
  ];
  \node [circle, scale=0.2, fill=black] at (3,240) {};
  \node [circle, scale=0.2, fill=black] at (4,200) {};
  \node [circle, scale=0.2, fill=black] at (9,250) {};
  \node [circle, scale=0.2, fill=black] at (11,270) {};
  \node [circle, scale=0.2, fill=black] at (16,235) {};
  \node [circle, scale=0.2, fill=black] at (17,255) {};
  \node [circle, scale=0.2, fill=black] at (23,270) {};
  \node [circle, scale=0.2, fill=black] at (25,300) {};
  \node [circle, scale=0.2, fill=black] at (26,320) {};
  \node [circle, scale=0.2, fill=black] at (27,325) {};
  \node [circle, scale=0.2, fill=black] at (29,295) {};
  \node [circle, scale=0.2, fill=black] at (30,285) {};
  \node [circle, scale=0.2, fill=black] at (32,260) {};
  \node [circle, scale=0.2, fill=black] at (34,255) {};
  \node [circle, scale=0.2, fill=black] at (36,259) {};
  \node [circle, scale=0.2, fill=black] at (46,280) {};
  \node [circle, scale=0.2, fill=black] at (48,370) {};
  \node [circle, scale=0.2, fill=black] at (62,360) {};
  \node [circle, scale=0.2, fill=black] at (64,385) {};
  \node [circle, scale=0.2, fill=black] at (69.5,400) {};
  \node [circle, scale=0.2, fill=black] at (68,380) {};
                         2
  \node [circle, scale=0.2, fill=red] at (42,220) {};
  \node at (42,185) {\footnotesize $c_i$};
  
\end{axis}
\end{tikzpicture}
\caption{}
\end{subfigure}
\caption{An example of an $\epsilon$-support vector machine considering the influence of the configuration option $o_{\mathit{Num}_1}$ on the property $y$ with (a) the original configuration space and (b) the transformed configuration space using a predefined kernel function.}
\label{fig:svm}
\end{figure}

First, we focus on \textit{$\epsilon$-support vector regression (SVR)}, which aims at identifying a flat function in the transformed space by allowing a prediction error of less than $\epsilon$ for each configuration. 
For configurations with a prediction error larger than $\epsilon$, the transformation is penalized~\cite{basak2007support,vapnik1995nature,smola2004tutorial}.
This way, this technique aims at identifying a transformation that predicts performance of all configurations from the learning set with an error of less than $\epsilon$.
In Figure~\ref{fig:svm}a, we show a configuration space for which it is not possible to correctly consider the influence of the option $o_{\mathit{Num}_1}$ using a linear function.
Using SVR, the space is transformed by the use of a non-linear kernel, as shown in Figure~\ref{fig:svm}b.
The applied transformation is penalized only by the prediction error of the configuration $c$, denoted by a red dot, as it is larger than $\epsilon$.
To predict an unobserved configuration, the configuration is also transformed using the same kernel function and then predicted in the transformed space.


Second, we consider \textit{Kernel Ridge Regression (KRR)}.
Here, the idea of kernel transformation is combined with \emph{ridge regression}, which penalizes models with large coefficients~\cite{Vovk2013}.
Ridge regression is especially useful when numerical instabilities occur, when input values are correlated with each other~\cite{Cristianini:1999:ISV:345662}, or to minimize the risk of overfitting~\cite{Murphy:2012:MLP:2380985}.
To this end, a penalty term is added to the rating of an applied transformation.
Depending on the weight of this term, complex predictor functions are more penalized, while permitting a higher prediction error on the learning set.
As a consequence, a large penalty term leads to simple models with possibly high error rates, while a too small penalty term might lead to overfitted models and a high generalization error.
As said previously, we account for such hyper-parameters in our study.


\subsection{k-Nearest-Neighbors Regression}

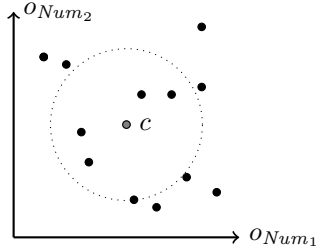
\begin{figure}
\centering
\begin{tikzpicture}
    \draw[->,thick] (0,0)--(3,0) node[right]{$o_{\mathit{Num}_1}$};
    \draw[->,thick] (0,0)--(0,3) node[right]{$o_{\mathit{Num}_2}$};

    \draw[fill]  (1,1) circle (0.05);
    \draw[fill]  (1.6,0.5) circle (0.05);
    \draw[fill]  (2.1,1.9) circle (0.05);
    \draw[fill]  (0.4,2.4) circle (0.05);
    \draw[fill]  (0.4,2.4) circle (0.05);
    \draw[fill]  (0.7,2.3) circle (0.05);
    \draw[fill]  (0.9,1.4) circle (0.05);
    \draw[fill]  (1.7,1.9) circle (0.05);
    \draw[fill]  (1.9,0.4) circle (0.05);
    \draw[fill]  (2.3,0.8) circle (0.05);
    \draw[fill]  (2.7,0.6) circle (0.05);
    \draw[fill]  (2.5,2.8) circle (0.05);
    \draw[fill]  (2.5,2) circle (0.05);
    
    \draw[fill,gray]  (1.5,1.5) circle (0.05);
    \draw  (1.5,1.5) circle (0.05);
    \node (interaction) at (1.75,1.5) {$c$};
    \draw[dotted]  (1.5,1.5) circle (1.01);
\end{tikzpicture}
\caption{An example for the k nearest-neighbors regression on a configuration space of two configuration options $o_{\mathit{Num}_1}$ and $o_{\mathit{Num}_2}$.}
\label{fig:KNN}
\end{figure}

In \textit{k-nearest-neighbors regression (kNN)}, configurations correspond to points in an n-dimensional vector space, where each dimension of the space corresponds to one configuration option~\cite{Mitchell:1997:ML:541177}, which is equivalent to other learning techniques, such as SVR.
In contrast to the other machine-learning techniques, kNN does not learn an underlying prediction model.
Instead, it considers only already measured neighboring configurations to produce a prediction.
To predict the performance of a configuration, kNN uses the measured performance values of the \textit{k}-closest configurations from the learning set to the configuration of interest~\cite{Mitchell:1997:ML:541177}.
In Figure~\ref{fig:KNN}, we can predict the performance of configuration $c$ by taking, for example, the mean value of the 5 configurations that lay inside the circle.
Different distance measures can be used, which we also consider during hyper-parameter optimization.
In general, this method is beneficial when we have no assumptions about the shape of the regression function to be learned~\cite{doi:10.1080/00031305.1992.10475879}.

\section{Sampling Strategies}\label{sec:Sampling}

A sampling strategy selects a set $L = \{c_1,...,c_n\}$ of configurations from the overall configuration space $C$, such that we can efficiently learn an accurate predictor $\Pi$.
Depending on the sampling strategy, different configurations are selected for the learning set.
For instance, some strategies aim at selecting those configurations that contain as many potential interactions as possible, whereas others aim at achieving a uniform coverage of the configuration space.

In this section, we present the most prominent sampling strategies that have been used for learning performance models in the literature~\cite{Guo2013, SGA+15, Medeiros:2016:CSA:2884781.2884793}. 
We differentiate between sampling strategies for binary and numeric options.
In Table~\ref{tab:OverviewOfTheStructredSamplingStrategiesIncludingTheNumberOfSelectedConfigurationOptions}, we provide an overview of the sampling strategies we consider in our study and the number of configurations they select.

\subsection{Sampling Binary Configuration Spaces}\label{sec:binaryHeuristics}

We selected 4 sampling strategies, 3 of which rely on a coverage criterion and one random sampling strategy.

\subsubsection{Option-wise strategy} \label{sec:OWheuristic}

The \textit{Option-wise strategy (OW)} aims at selecting a learning set, in which each option is selected at least once in a configuration and, simultaneously, the number of other selected options is minimized to reduce the influence of unknown interactions.
Based on this set of configurations, the individual influences of all binary configuration options can be learned.
The size of the learning set is linear in the number of binary options of the configurable system.
This sampling strategy has been used by Medeiros et al.~\cite{Medeiros:2016:CSA:2884781.2884793} and Apel et al.~\cite{ARW+13}.

\subsubsection{T-wise strategy}

The \textit{T-wise strategy} aims at selecting a learning set, in which each T-tuple of options is selected, at least, once (T $ > $ 1).
As for the OW strategy, the number of other combinations of options is minimized.
For example, the 2-wise strategy selects one configuration per pair of binary configuration options, while the 3-wise heuristics selects configurations for each triple of binary configuration options.
The number of the selected configurations is $|O_{Bin}|^T$
In our experiments, we consider $T = 2$ and $T = 3$, which we call henceforth T2 and T3. 
This sampling strategy has been used, for example, by Medeiros et al.~\cite{Medeiros:2016:CSA:2884781.2884793}, Perrouin et al.~\cite{Perrouin:2010:AST:1828417.1828490}, and Nie and Leung~\cite{Nie:2011:SCT:1883612.1883618}.

\subsubsection{Negative option-wise strategy}

The \textit{Negative Option-wise strategy (NegOW)} selects one configuration for each option, in which the option being considered is deselected and all other options are selected.
In addition, one configuration in which all options are enabled (i.e., all-yes configuration) is added to the learning set.
Based on these configurations, the influence of all options in the presence of all interactions can be identified.
The number of configurations selected is linear in the number of configuration options of the system.
This sampling strategy has been used, for example, by Siegmund et al.~\cite{SGA+15} or by Medeiros et al.~\cite{Medeiros:2016:CSA:2884781.2884793}.

\subsubsection{Random}

We also apply \textit{Random sampling (RB)} on the binary configuration space, which is a non-trivial task in the presence of constraints among configuration options~\cite{Chakraborty:2014:DSW:2892753.2892792}.
Thus, we first generate all valid binary configurations and then select a pseudo random distributed subset using a specific seed.
To obtain comparable results, we choose the same number of configurations for random sampling as being selected by the OW, T2, and T3 strategies, respectively.
We abbreviate a random sampling strategy with RB(0,T2) if it selects a number of configurations equal to T2 and uses a random seed of 0.
When considering the average prediction accuracy using this size and different seed, we abbreviate this by omitting the seed RB(\_,T2).
This sampling strategy has been used, for example, by Guo et al.~\cite{Guo2013}, Temple et al.~\cite{temple:hal-01467299}, and Medeiros et al.~\cite{Medeiros:2016:CSA:2884781.2884793}.

\subsection{Sampling Numeric Configuration Spaces}\label{sec:DoE}

To sample a numeric configuration space, we consider a set of strategies known under the umbrella of \emph{design of experiments}, which have been originally developed as screening techniques to purposefully select configurations to learn a response function~\cite{myers2009response}.
In total, we consider 6 sampling strategies, which make different assumptions about the influence of configuration options on performance and, thus, the expected structure of the response function.
In addition, we selected random sampling because it is one prominent and frequently used strategy in the literature.

\begin{figure}
\begin{subfigure}[b]{.3\linewidth}
\begin{tikzpicture}

    \def\side{1.5}
    \def\ptSize{1.5pt}

    \filldraw[white] (\side/2,0,\side) circle (\ptSize);

    \draw (\side,0,0) -- (\side,\side,0) -- (0,\side,0);
    \draw (0,0,\side) -- (\side,0,\side) -- (\side,\side,\side) -- (0,\side,\side) -- (0,0,\side);
    \draw (\side,0,0) -- (\side,0,\side) ;
    \draw (\side,\side,0) -- (\side,\side,\side);
    \draw (0,\side,0) -- (0,\side,\side);

    \draw[dashed]  (0,0,\side) -- (0,0,0);
    \draw[dashed]  (\side,0,0) -- (0,0,0);
    \draw[dashed]  (0,\side,0) -- (0,0,0);

    \filldraw (\side/2,\side/2,\side/2) circle (\ptSize);
    
    \filldraw (0,\side/2,\side/2) circle (\ptSize);
    \filldraw (\side/4,\side/2,\side/2) circle (\ptSize);
    \filldraw (3*\side/4,\side/2,\side/2) circle (\ptSize);
    \filldraw (4*\side/4,\side/2,\side/2) circle (\ptSize);

    \filldraw (\side/2,0,\side/2) circle (\ptSize);
    \filldraw (\side/2,\side/4,\side/2) circle (\ptSize);
    \filldraw (\side/2,3*\side/4,\side/2) circle (\ptSize);
    \filldraw (\side/2,4*\side/4,\side/2) circle (\ptSize);

    \filldraw (\side/2,\side/2,0) circle (\ptSize);
    \filldraw (\side/2,\side/2,\side/4) circle (\ptSize);
    \filldraw (\side/2,\side/2,3*\side/4) circle (\ptSize);
    \filldraw (\side/2,\side/2,4*\side/4) circle (\ptSize);
    
    \draw[densely dotted]  (0,\side/2,\side/2) -- (\side,\side/2,\side/2);
    \draw[densely dotted]  (\side/2,0,\side/2) -- (\side/2,\side,\side/2);
    \draw[densely dotted]  (\side/2,\side/2,0) -- (\side/2,\side/2,\side);

\end{tikzpicture}
\caption{}
\end{subfigure}
\begin{subfigure}[b]{.3\linewidth}
\begin{tikzpicture}

    \def\side{1.5}
    \def\ptSize{1.5pt}

    \draw (\side,0,0) -- (\side,\side,0) -- (0,\side,0);
    \draw (0,0,\side) -- (\side,0,\side) -- (\side,\side,\side) -- (0,\side,\side) -- (0,0,\side);
    \draw (\side,0,0) -- (\side,0,\side) ;
    \draw (\side,\side,0) -- (\side,\side,\side);
    \draw (0,\side,0) -- (0,\side,\side);

    \draw[dashed]  (0,0,\side) -- (0,0,0);
    \draw[dashed]  (\side,0,0) -- (0,0,0);
    \draw[dashed]  (0,\side,0) -- (0,0,0);

    \filldraw (\side/2,\side/2,\side/2) circle (\ptSize);
    
    \filldraw (\side/2,\side,\side) circle (\ptSize);
    \filldraw (\side/2,0,\side) circle (\ptSize);
    \filldraw (\side/2,\side,0) circle (\ptSize);
    \filldraw (\side/2,0,0) circle (\ptSize);

    \filldraw (\side,\side/2,\side) circle (\ptSize);
    \filldraw (0,\side/2,\side) circle (\ptSize);
    \filldraw (\side,\side/2,0) circle (\ptSize);
    \filldraw (0,\side/2,0) circle (\ptSize);
    
    \filldraw (\side,\side,\side/2) circle (\ptSize);
    \filldraw (0,\side,\side/2) circle (\ptSize);
    \filldraw (\side,0,\side/2) circle (\ptSize);
    \filldraw (0,0,\side/2) circle (\ptSize);

    \draw[densely dotted]  (0,0,\side/2) -- (\side,0,\side/2);
    \draw[densely dotted]  (0,\side,\side/2) -- (\side,\side,\side/2);
    \draw[densely dotted]  (0,0,\side/2) -- (0,\side,\side/2);
    \draw[densely dotted]  (\side,0,\side/2) -- (\side,\side,\side/2);
    
    \draw[densely dotted]  (\side/2,0,0) -- (\side/2,0,\side);
    \draw[densely dotted]  (\side/2,\side,0) -- (\side/2,\side,\side);
    \draw[densely dotted]  (\side/2,\side,0) -- (\side/2,0,0);
    \draw[densely dotted]  (\side/2,\side,\side) -- (\side/2,0,\side);
    
    \draw[densely dotted]  (\side,\side/2,\side) -- (0,\side/2,\side);
    \draw[densely dotted]  (\side,\side/2,0) -- (\side,\side/2,\side);
    \draw[densely dotted]  (\side,\side/2,0) -- (0,\side/2,0);
    \draw[densely dotted]  (0,\side/2,\side) -- (0,\side/2,0);

    \draw[densely dotted]  (0,\side/2,\side/2) -- (\side,\side/2,\side/2);
    \draw[densely dotted]  (\side/2,0,\side/2) -- (\side/2,\side,\side/2);
    \draw[densely dotted]  (\side/2,\side/2,0) -- (\side/2,\side/2,\side);

\end{tikzpicture}
\caption{}
\end{subfigure}
\begin{subfigure}[b]{.3\linewidth}
\begin{tikzpicture}
    \def\side{1.5}
    \def\ptSize{1.5pt}
    \def\starUp{0.75}
    \def\starDo{0.25}
    
    \filldraw[white] (\side/2,0,\side) circle (\ptSize);
    
    \draw (\side,0,0) -- (\side,\side,0) -- (0,\side,0);
    \draw (0,0,\side) -- (\side,0,\side) -- (\side,\side,\side) -- (0,\side,\side) -- (0,0,\side);
    \draw (\side,0,0) -- (\side,0,\side) ;
    \draw (\side,\side,0) -- (\side,\side,\side);
    \draw (0,\side,0) -- (0,\side,\side);

    \draw[dashed]  (0,0,\side) -- (0,0,0);
    \draw[dashed]  (\side,0,0) -- (0,0,0);
    \draw[dashed]  (0,\side,0) -- (0,0,0);

    \filldraw (\side/2,\side/2,\side/2) circle (\ptSize);
    \filldraw (0,\side/2,\side/2) circle (\ptSize);
    \filldraw (\side,\side/2,\side/2) circle (\ptSize);
    \filldraw (\side/2,0,\side/2) circle (\ptSize);
    \filldraw (\side/2,\side,\side/2) circle (\ptSize);
    \filldraw (\side/2,\side/2,0) circle (\ptSize);
    \filldraw (\side/2,\side/2,\side) circle (\ptSize);

    \filldraw (\starUp*\side,\starUp*\side,\starUp*\side) circle (\ptSize);
    
    \filldraw (\starDo*\side,\starUp*\side,\starUp*\side) circle (\ptSize);
    \filldraw (\starUp*\side,\starDo*\side,\starUp*\side) circle (\ptSize);
    \filldraw (\starUp*\side,\starUp*\side,\starDo*\side) circle (\ptSize);
    
    \filldraw (\starDo*\side,\starDo*\side,\starUp*\side) circle (\ptSize);
    \filldraw (\starDo*\side,\starUp*\side,\starDo*\side) circle (\ptSize);
    \filldraw (\starUp*\side,\starDo*\side,\starDo*\side) circle (\ptSize);
    
    \filldraw (\starDo*\side,\starDo*\side,\starDo*\side) circle (\ptSize);
    
     \draw[densely dotted] (\side/2,\side/2,\side/2) circle (\side/2);
    
     \draw[densely dotted] (0,\side/2,\side/2) arc (180:360:0.75 and 0.3);
     \draw[densely dotted] (\side,\side/2,\side/2) arc (0:180:0.75 and 0.3);
     \draw[densely dotted] (\side/2,0,\side/2) arc (270:90:0.3 and 0.75);
     \draw[densely dotted] (\side/2,\side,\side/2) arc (90:270:-0.3 and 0.75);
    
\end{tikzpicture}
\caption{}
\end{subfigure}
\caption{Selected configurations for different numeric sampling strategies on a configuration space of three options: (a) One Factor At A Time; (b) Box-Behnken Design; (c) Central Composite Inscribed Design}
\label{fig:expDesign1}
\end{figure}
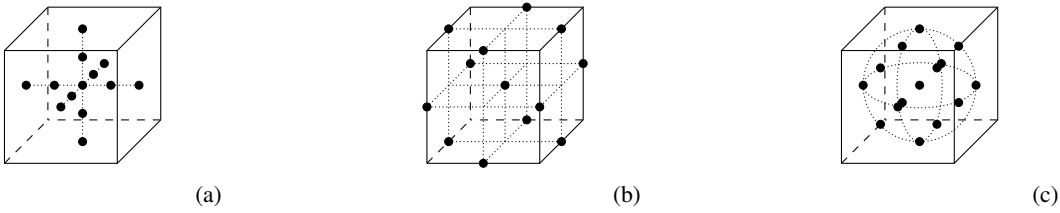

\begin{table*}[t]
  \caption{Overview of sampling strategies including the abbreviation we use throughout the paper and the number of selected configurations.}
  \label{tab:OverviewOfTheStructredSamplingStrategiesIncludingTheNumberOfSelectedConfigurationOptions}
  \centering
  \begin{tabular}{llc}
    \toprule
    Sampling strategy & Abbrv. & \#configurations \\
    \midrule
    Option-wise heuristics & OW & $|O_{\mathit{Bin}}|$ \\
    T-wise heuristics & Tt (e.g., T1, T2, T3) & $|O_{\mathit{Bin}}|^t$ \\
    Negative option-wise heuristics & NegOW & $|O_{\mathit{Bin}}|$ \\
    \vspace{0.4em}
    Random (Binary) & RB & user defined\\
    One Factor At A Time & OFAT & $|O_{\mathit{Num}}| \times ($user defined$-1) + 1$ \\
    Box-Behnken Design & BBD & $2 \times |O_{\mathit{Num}}| \times (|O_{\mathit{Num}}|-1) + 1$ \\
    Central Composite Inscribed Design & CCI & $2^{|O_{\mathit{Num}}|} + 2 \times |O_{\mathit{Num}}| + 1$  \\
    Plackett-Burman Design & PBD & defined by seeds\\
    D-Optimal Design & DOD & user defined \\
    Random (Numeric) & RN & user defined \\
    \bottomrule
	\end{tabular}
\end{table*}
\normalsize

\subsubsection{One Factor At A Time}

The \textit{One Factor At A Time (OFAT)} design assumes that there are no interactions among numeric configurations options~\cite{10.2307/2685731}, which corresponds to the OW strategy for binary options (cf. Section~\ref{sec:OWheuristic}).
Thus, the learning set consists of a set of configurations for each option, for which the numeric value of the respective option is varied; all other options are left unchanged, usually kept at the center of their value domains.
An example for this design is given in Figure~\ref{fig:expDesign1}a, where each dimension of the cube represents the value domain of one option.
Using this design, the number of configurations selected is linear in the number of options being considered.
This sampling strategy has been used, for example, by Morris~\cite{morris1991factorial} and Campolongo et al.~\cite{campolongo2007effective}.

\subsubsection{Box-Behnken Design}

The \textit{Box-Behnken Design (BBD)} selects configurations such that one can derive a second-order response-surface model from the learning set~\cite{BoxBehnkenDesign, Montgomery:2006:DAE:1206386}.
Such a model considers quadratic influences for each numeric option as well as 2-wise interactions among configuration options.
BBD selects basically a subset of the configurations defined by the $3^k$ Full Factorial Design.
The $3^k$ Full Factorial Design creates and uses all value combination of $k$ configuration options, where 3 values of each option are selected (e.g., min, max, and center of the value range). 
When considering $|O_{\mathit{Num}}|$ configuration options, BBD selects $2 \times |O_{\mathit{Num}}|\times(|O_{\mathit{Num}}|-1) + 1$ different value combinations~\cite{ferreira2007box}.
It is often used when there is only little or no interest in predicting configurations on the extrema of the configuration space~\cite{Nguyen2008294}.
An example of this design for three options is given in Figure~\ref{fig:expDesign1}b.
This sampling strategy has been used, for example, by Aslan and Cebeci~\cite{ASLAN200790} and Annadurai and Sheeja~\cite{Annadurai1998}.
In our experiments, we refrain from considering the $3^k$ Full Factorial Design or the $2^k$ Full Factorial Design, because the number of configurations selected by these two designs is exponentially in the number of options being considered, while it is only possible to learn quadratic or linear influences of the options.

\subsubsection{Central Composite Design}

The \textit{Central Composite Design (CCD)} was introduced by Box and Wilson~\cite{BoxWilson1951}. 
It is one of the most important designs for deriving second-order response-surface models~\cite{myers2009response,Montgomery:2006:DAE:1206386}.
The configurations selected by CCD consists of three sets:
(I) all configurations considered by a $2^k$ Full Factorial Design\footnote{Similar to the $3^k$ Full Factorial Design, the $2^k$ Full Factorial Design considers all value combinations of options. However, only two instead of three values of each option are considered.};
(II) a set containing of $2\times|O_{\mathit{Num}}|$ configurations with a distance of $\alpha$ to the center of the normalized configuration space\footnote{In a normalized configuration space, the value domains of the numeric configuration options are mapped to [0,1].};
(III) a configuration with all numeric options on their center value.
These three sets lead to $2^{|O_{\mathit{Num}}|} + 2\times|O_{\mathit{Num}}| + 1$ different configurations for $|O_{\mathit{Num}}|$ numeric configuration options~\cite{ferreira2007box, KhuriMukho10}.

Depending on the value of $\alpha$, we differentiate between three types of CCDs (Central Composite Inscribed Design, Central Composite Face Centered Design, and Central Composite Circumscribed Design), where we decide to use the \textit{Central Composite Inscribed (CCI) Design} with $|\alpha| < 1$ in a normalized configuration space.
We give an example for the CCI design for three parameters in Figure~\ref{fig:expDesign1}c.
This sampling strategy has been used, for example, by Yeten et al.~\cite{yeten2005comparison} and Garg and Tai~\cite{garg2013selection}.

\subsubsection{Plackett-Burman Design}

Proposed by Plackett and Burman in 1946, the \textit{Plackett-Burman Design (PBD)} aims at identifying influences of configuration options in configuration spaces in which interactions are negligible compared to the individual influences of the configuration options~\cite{PlackettBurman1946}.
In contrast to other experimental designs, such as CCD, where the number of selected configurations depends on the number of configuration options, PBD provides a set of seeds to select from.
In the construction of the seeds, Galois fields are used to obtain a cyclic solution, where each level of an option (i.e., a specific value in the option's domain) is considered in the same number of configurations.
For more details, we refer to the work of Plackett and Burman~\cite{PlackettBurman1946}.
The seeds define the number of selected configurations ($\#c$) and the number of different values ($\#v$) that are considered in the sampling process.
We use the seeds proposed by Plackett and Burman, which are available on the supplementary Web site.
These seeds are designed to maximize the number of individual influences and lower-order interactions, which can be identified from the configurations.
For example, the first seed defines that 9 different configurations are selected and 3 different values for each option are considered in the sampling process.
In what follows, we denote this seed with PBD(9,3).
Overall, we consider four different seeds in our experiments leading us to four different versions of PBD: PBD(9,3), PBD(25,5), PBD(49,7), and PBD(125,5).

\begin{figure}
	\centering
		\includegraphics{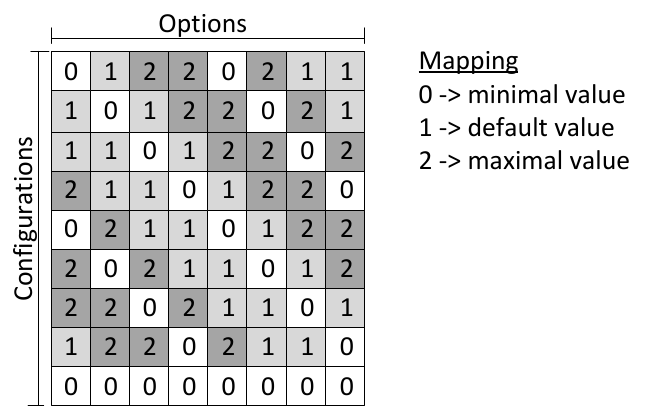}
	\caption{Configurations selected by the Plackett-Burman design using seed (9,3): 9 configurations and 3 domain values.}
	\label{fig:PlackettBurman}
\end{figure}

Based on these seeds, the configurations of the learning set are selected.
To this end, the seed is used to define the first configuration.
The later configurations are defined by shifting the seed to the right, which is illustrated in Figure~\ref{fig:PlackettBurman}.
Subsequently, the values used in the seed are mapped to actual values of the value domain of the configuration options.
Here, we apply an equidistant mapping from the minimal to maximal value of the value domains of the numeric options.
This sampling strategy has been used, for example, by Yeten et al.~\cite{yeten2005comparison}, Jacques et al.~\cite{Jacques1999}, and Krishnan et al.~\cite{Krishnan1998}.


\subsubsection{Optimal Designs}

Optimal Designs select configurations based on different mathematical criteria.
This stands in contrast to the previously mentioned experimental designs, where the configurations are selected based on specific structural patterns of the configuration space.
In our work, we focus on the \textit{D-Optimal Design (DOD)}.
This design aims at minimizing the determinant of the dispersion matrix $(X^TX)^{-1}$, where all configurations selected by this design are stored in the model matrix $X$~\cite{Smith1918}.
This minimization of the determinant of the dispersion matrix leads to configurations having the largest possible distance to each other; hence, covering the whole configuration space as equal as possible.

However, selecting an optimal set of configurations is a non-trivial task.
It does not scale because it would require to enumerate all possible configurations.
From this set of configurations, a subset is selected and iteratively refined to be optimal regarding the optimization criteria by replacing one or more configurations from this set with other configurations, being not in the set.
For this purpose, there exist different algorithms, such as the Fedorov algorithm~\cite{fedorov1972theory}, the DETMAX algorithm~\cite{mitchellExchangeAlgorithm}, or the k-exchange algorithm~\cite{kexchange}.
The replacement procedure is performed until replacing one configuration of the model matrix does not lead to benefits regarding the optimization criterion.
However, it might happen that the algorithms are not able to identify the optimal set of configuration but only an almost optimal set.
In our experiments, we use the k-exchange algorithm to select 50 and 125 configurations, denoted as DOD(50) and DOD(125), being comparable with PBD(49,9) and PBD(125,5).
This sampling strategy has been used, for example, by Yeten et al.~\cite{yeten2005comparison} and Fang and Perera~\cite{FANG2011717}.

\subsubsection{Random Design}

We also consider a \textit{Random Design (RN)} for the numeric configuration spaces in our experiments.
Although this sampling strategy has some drawbacks, it is often used in the literature~\cite{temple:hal-01467299, Ganapathi:2009:CML:1855591.1855592, bergstra2012machine}.
One drawback, for example, is that the randomness in the selection process might lead to a non-uniformly, clustered learning set with respect to the overall configuration space~\cite{saltelli2008global}.
To increase reproducibility, we apply a pseudo random sampling using a predefined seed.
This sampling strategy has been used, for example, by Temple et al.~\cite{temple:hal-01467299}, Ganapathi et al.~\cite{Ganapathi:2009:CML:1855591.1855592}, and Bergstra et al.~\cite{bergstra2012machine}.
In our experiments, we select 50 and 125 configurations, RN(50), RN(125), to be able to compare results with DOD and PBD of the respective size.

\section{Methodology}\label{sec:Meth}

In this section, we discuss the methodology of our study.
In particular, we describe our approach of hyper-parameter optimization of the machine-learning techniques, the independent and dependent variables, the subject systems, and the study setup.

\subsection{Hyper-Parameter Optimization} \label{sec:mlParamOpt}

All machine-learning techniques that we consider offer parameters, called \textit{hyper parameters}, which affect the efficiency of learning and the accuracy of predictions.
To make a fair comparison of the different machine-learning techniques with respect to their prediction accuracies, it is necessary to optimize their hyper parameters~\cite{DBLP:journals/infsof/FuMS16}.
As a consequence, we perform a hyper-parameter optimization of the machine-learning techniques, using k-fold cross validation, prior to the learning procedure itself.
As the choice of the sampling set is relevant for optimization, we performed hyper-parameter optimization for each machine-learning techniques based on a specific combination of a binary and a numeric sampling strategy.
In Table~\ref{tab:mlOptimization}, we show the parameters that we incorporated in the optimization.
For optimization, we perform a random search, which is considered to be more efficient than performing a full grid search~\cite{Bergstra:2012:RSH:2188385.2188395}.
For reproducibility, we provide the hyper-parameter values being identified as optimal on the supplementary Web site.

\begin{table*}[t]
  \caption{Hyper parameters of the machine-learning techniques being considered in the parameter optimization.}
  \label{tab:mlOptimization}
  \centering
  \begin{tabular}{llp{7.5cm}}
    \toprule
    Learning technique & Parameters & Description\\
    \midrule
    SVR & C & Penalty parameter of the error term. \\
        & epsilon & Defines the size of the $\epsilon$-tube.\\
        & coef0 & Independent term in the kernel function.\\
        & shrinking & Parameter to speed up the optimization.\\
        & tol & Tolerance for the stopping criterion.\\
                
    CART & splitter & Defines the strategy used to choose a split in each node.\\
         & max\_features & Defines the number of configuration options being considered when identifying the best split.\\
         & min\_samples\_leaf & The minimal number of configurations required in each leaf node. \\
         & random\_state & The seed that is used in the random number generator.\\
    
    RF & n\_estimators & The number of trees in the forest.\\
       & max\_features & Defines the number of configuration options considered when identifying the best split.\\
       & random\_state & The seed that is used in the random number generator. \\
    
    kNN & n\_neighbors & Number of configurations considered in a prediction.\\
        & weights & Weights of the neighbor configurations.\\
        & algorithm & Algorithm used to compute the neighbors.\\
        & p & Parameter that defined the power of the Minkowski metric (e.g., 2 for the Euclidean metric).\\
    
    KRR & alpha & Parameter that aims at reducing the variance of the predictions.\\
        & kernel & Defines the kind of kernel being considered (e.g., linear).\\
        & degree & The degree of the polynomial kernel.\\
        & gamma & Parameter for the radial basis function used in the kernel.\\
    
    MR & minImprovement & Minimal improvement per round that is used as stopping criterion.\\
       & lossFunction & The loss function used in the regression.\\
       & functionTypes & Defines whether complex interactions such as $\frac{o_{\mathit{Num_1}}}{o_{\mathit{Num_2}}}$ are considered in the learning.\\
    \bottomrule
  \end{tabular}
\end{table*}


\subsection{Experiment Variables}

Next, we present the dependent and independent variables of our study, and we describe how we use them to answer our research questions.

\begin{figure}
\begin{tikzpicture}
\small
  \def\side{1.7}
  
  \def\first{.5}
  \def\dist{6.5}
  \def\ptSize{1.3pt}

  \draw (\side+\first,0,0) -- (\side+\first,\side,0) -- (0+\first,\side,0);
  \draw (0+\first,0,\side) -- (\side+\first,0,\side) -- (\side+\first,\side,\side) -- (0+\first,\side,\side) -- (0+\first,0,\side);
  \draw (\side+\first,0,0) -- (\side+\first,0,\side) ;
  \draw (\side+\first,\side,0) -- (\side+\first,\side,\side);
  \draw (0+\first,\side,0) -- (0+\first,\side,\side);

  \draw[dashed]  (0+\first,0,\side) -- (0+\first,0,0);
  \draw[dashed]  (\side+\first,0,0) -- (0+\first,0,0);
  \draw[dashed]  (0+\first,\side,0) -- (0+\first,0,0);

  \node (interaction) at (0+0.2+\first, -\side/2-0.1, 0) {\textsf{machine-learning}};
  \node (interaction) at (0+0.2+\first, -\side/2-0.1-0.35, 0) {\textsf{technique}};

  \node [anchor=west] (interaction) at (\side+\first, \side/2+0.35, 0) {\textsf{numeric sampling}};
  \node [anchor=west] (interaction) at (\side+\first, \side/2, 0) {\textsf{strategy}};
  
  \node [anchor=west] (interaction) at (\side+\first+0.15, 0, \side/2) {\textsf{binary sampling}};
  \node [anchor=west] (interaction) at (\side+\first+0.15, 0-0.35, \side/2) {\textsf{strategy}};
  
  \node [anchor=south] (interaction) at (\side/2+\first, \side, 0) {\textsf{subject system $1$}};
  
  
   \node [anchor=west] (interaction) at (\side/2+\dist/2, \side/3, 0) {...};
  

  \draw (\side+\dist,0,0) -- (\side+\dist,\side,0) -- (0+\dist,\side,0);
  \draw (0+\dist,0,\side) -- (\side+\dist,0,\side) -- (\side+\dist,\side,\side) -- (0+\dist,\side,\side) -- (0+\dist,0,\side);
  \draw (\side+\dist,0,0) -- (\side+\dist,0,\side) ;
  \draw (\side+\dist,\side,0) -- (\side+\dist,\side,\side);
  \draw (0+\dist,\side,0) -- (0+\dist,\side,\side);

  \draw[dashed]  (0+\dist,0,\side) -- (0+\dist,0,0);
  \draw[dashed]  (\side+\dist,0,0) -- (0+\dist,0,0);
  \draw[dashed]  (0+\dist,\side,0) -- (0+\dist,0,0);

  \node (interaction) at (0+0.2+\dist, -\side/2-0.1, 0) {\textsf{machine-learning}};
  \node (interaction) at (0+0.2+\dist, -\side/2-0.1-0.35, 0) {\textsf{technique}};
  
  \node [anchor=west] (interaction) at (\side+\dist, \side/2+0.35, 0) {\textsf{numeric sampling}};
  \node [anchor=west] (interaction) at (\side+\dist, \side/2, 0) {\textsf{strategy}};
  
  \node [anchor=west] (interaction) at (\side+\dist+0.15, 0, \side/2) {\textsf{binary sampling}};
  \node [anchor=west] (interaction) at (\side+\dist+0.15, 0-0.35, \side/2) {\textsf{strategy}};
    
  \node [anchor=south] (interaction) at (\side/2+\dist, \side, 0) {\textsf{subject system $n$}};
  
  \normalsize
\end{tikzpicture}
\caption{Study design.}
\label{fig:studySetup}
\end{figure}
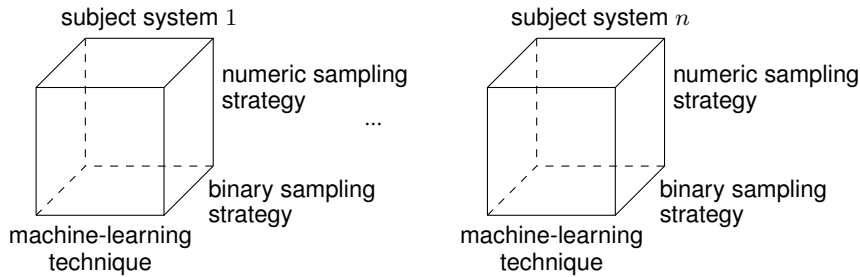

\textit{Independent variables:}
The independent variables of our study are (I) the choice of the machine-learning technique $l \in L$ with $L = \{\textsf{CART, kNN, KRR, MR, RF, SVR}\}$, (II) the choice of the binary sampling strategy $b \in B$ with $B = \{\textsf{OW, NegOW, T2,}$ \newline$\textsf{ T3, RB(\_,OW), RB(\_,T2), RB(\_,T3)}\}$, (III) the choice of the numeric sampling strategy $n \in N$ with $N = \{\textsf{OFAT, BBD, CCI,}$ \newline$\textsf{ PBD(9,3), PBD(25,5), PBD(125,5), PBD(49,7), DOD(50), DOD(125), RN(50), RN(125)}\}$, and (IV) the subject system $s \in S$.
An overview of the independent variables is provided in Figure~\ref{fig:studySetup}.
To answer our research questions, we performed one experiment for each of the machine-learning techniques using each combination of a binary and a numeric sampling strategy on each of the subject systems.

\textit{Dependent variables:}
The dependent variable in our study is the average error rate of the performance predictions for all configurations of a given subject system $s$, which we define as 

\begin{center}
$\overline{e}(l,b,n,s) = \frac{ \sum_{c \in C(s)} \frac{|\textrm{measured}(c) - \Pi(c)|}{\textrm{measured}(c)} }{|C(s)|} $
\end{center}

\noindent where $\textrm{measured}(c)$ is the measured performance of the configuration $c \in C(s)$ and $\Pi(c)$ is the predicted performance value of this configuration, which depends on the selected machine-learning technique $l \in L$, the binary sampling strategy $b \in B$, the numeric sampling strategy $n \in N$, and the subject system $s \in S$.
We use this dependent variable as a measure to answer our research questions, by comparing the error rates achieved by different combinations of machine-learning techniques and sampling strategies respectively.

For answering RQ1.1, we determine whether there is a machine-learning technique that leads to smaller error rates compared to all other considered machine-learning techniques when using the same combination of a binary and a numeric sampling strategy on the same subject system:

\begin{center}
$\exists \, l_i \in L : \, \mathop{\forall}_{\substack{b \in B, n \in N, s \in S, l_j \in L, l_j \neq l_i}} \overline{e}(l_i,b,n,s) < \overline{e}(l_j,b,n,s)$.
\end{center}

In RQ1.2, we are interested in the stability of the machine-learning techniques:

\begin{center}
$\exists \, l_i \in L : \mathop{\forall}_{\substack{s \in S, l_j \in L, l_j \neq l_i}} ( \overline{e}_{max}(l_i,s) - \overline{e}_{min}(l_i,s)) < ( \overline{e}_{max}(l_j,s) - \overline{e}_{min}(l_j,s))$.
\end{center}

\noindent Here, we consider a machine-learning technique to be more stable, when the mean error rates of it's predictions are less affected by the change of the sampling strategy.
To this end, we determine the maximal mean error rate $\overline{e}_{max}(l,s) = \displaystyle\mathop{max}_{\substack{b \in B, n \in N}}(\overline{e}(l,b,n,s))$ and the minimal mean error rate $\overline{e}_{min}(l,s) = \displaystyle\mathop{min}_{\substack{b \in B, n \in N}}(\overline{e}(l,b,n,s))$ of a machine learning technique $l$ for a subject system $s$.

Much like RQ1.1, we answer RQ2.1 by determining whether there is a binary or a numeric sampling strategy that leads to the smallest error rates compared to any other binary or numeric sampling strategy respectively, when using the same machine-learning technique and the same numeric sampling or binary sampling strategy on the same subject system:

\begin{center}
$\exists \, b_i \in B : \, \mathop{\forall}_{\substack{l \in L, n \in N, s \in S, b_j \in B, b_j \neq b_i}}  \overline{e}(l,b_i,n,s) < \overline{e}(l,b_j,n,s)$, and

$\exists \, n_i \in N : \, \mathop{\forall}_{\substack{l \in L, b \in B, s \in S, n_j \in N, n_j \neq n_i}} \overline{e}(l,b,n_i,s) < \overline{e}(l,b,n_j,s)$.
\end{center}

To answer RQ2.2, we determine whether there is a binary or a numeric sampling strategy, when changing the machine-learning technique has the smallest impact on the error rates of all binary or numeric sampling strategies.
For binary sampling strategies, we have:

\begin{center}
$\exists \, b_i \in B : \mathop{\forall}_{\substack{s \in S, b_j \in B, b_j \neq b_i}} ( \overline{e}_{max}(b_i,s) - \overline{e}_{min}(b_i,s)) < ( \overline{e}_{max}(b_j,s) - \overline{e}_{min}(b_j,s))$.
\end{center}

Here, the maximal mean error rate of a binary sampling strategy for a subject system is defined as $ \overline{e}_{max}(b,s) = \displaystyle\mathop{max}_{\substack{l \in L, n \in N}}\overline{e}(l,b,n,s)$, and the minimal mean error rate as $ \overline{e}_{min}(b,s) = \displaystyle\mathop{min}_{\substack{l \in L, n \in N}}\overline{e}(l,b,n,s)$.

For numeric sampling strategies, we have:

\begin{center}
$\exists \, n_i \in N : \mathop{\forall}_{\substack{s \in S, n_j \in N, n_j \neq n_i}} ( \overline{e}_{max}(n_i,s) - \overline{e}_{min}(n_i,s)) < ( \overline{e}_{max}(n_j,s) - \overline{e}_{min}(n_j,s))$.
\end{center}

The maximal mean error rate of a numeric sampling strategy for a subject system is determined as $ \overline{e}_{max}(n,s) = \displaystyle\mathop{max}_{\substack{l \in L, b \in B}} \overline{e}(l,b,n,s)$, and the minimal mean error rate as $\overline{e}_{min}(n,s) = \displaystyle\mathop{min}_{\substack{l \in L, b \in B}} \overline{e}(l,b,n,s)$.

For RQ3.1, where we are interested in whether there is a single combination of a machine-learning technique, a numeric sampling strategy, and a binary sampling strategy leading to smaller error rates as compared to all other combinations considering the same subject system:

\begin{center}
$\exists \, l_i \in L, b_j \in B, n_k \in N : \, \mathop{\forall}_{\substack{l_x \in L, b_y \in B, n_z \in N, s \in S, l_i \neq l_x \vee b_j \neq b_y \vee n_k \neq n_z}} \overline{e}(l_i,b_j,n_k,s) < \overline{e}(l_x,b_y,n_z,s)$.
\end{center}

Note that some of the sampling strategies select different numbers of configurations for the learning set, which likely also affect the accuracy of the learned model.
We will account for this variation when comparing prediction accuracies by presenting and discussing the tradeoff between size of the learning set and prediction accuracy.
 
\subsection{Subject Systems}\label{sec:subSystems}

In our experiments, we consider 6 real-world configurable software systems that originate from different performance-critical domains and that have been implemented in different programming languages.
In Table~\ref{table:caseStudies}, we provide an overview of their configuration spaces, including the number of binary and numeric configuration options, the number of constraints, the number of configurations, and the performance measure of the system that we aim at predicting.
For different systems, we selected different performance measures because the whole approach of using machine-learning techniques is not tailored to a specific non-functional property, in general. 
In what follows, we provide a brief description of each subject system:

\begin{itemize}
  
  \item \textsc{DUNE MGS} is a multigrid implementation based on the DUNE framework\footnote{https://www.dune-project.org/}.
  The DUNE framework offers a large number of concepts and algorithms that can be used to numerically solve partial differential equations.
  DUNE MGS offers the possibility of selecting one of two preconditioners and one of four solvers.
  Furthermore, the number of pre-smoothing and post-smoothing steps can be defined.
  In our experiments, we aim at strong scaling, where we increased the size of the computation domain and keep the computational power constant.
  This variability yields 2\,304 configurations.
  In our experiments, we measured the time to solve Poisson's equation on a Dell OptiPlex-9020 with an Intel i5-4570 Quad Code and 32 GB RAM (Ubuntu 13.4).

  \item 
  \textsc{gemm} is a matrix multiplication code from the test suite of Polly.
  Polly is a high-level loop and data-locality optimization infrastructure for LLVM\footnote{https://polly.llvm.org/}.
  We consider configuration options that define, for example, the vectorizer that is used in the compilation process and the data dependency analysis.
  In our experiments, we considered 60\,959 different configurations.
  For each of the configurations, we measured the execution time of \textit{gemm} using the extra large dataset of its benchmark suite\footnote{http://web.cse.ohio-state.edu/~pouchet.2/software/polybench/}.

  \item \textsc{HSMGP} is a scalable multigrid solver that is implemented to run on large-scale systems, such as the JuQueen supercomputer in J{\"u}lich\footnote{http://www.fz-juelich.de/ias/jsc/EN/Expertise/Supercomputers/JUQUEEN/JUQUEEN\_node.html}. 
  In our experiments, we consider a set of different smoothers and coarse-grid solvers as well as a variable number of pre-smoothing and post-smoothing steps.
  We performed weak scaling experiments and took the number of nodes the computation run on into account.
  Overall, we measured 3\,456 different configurations.
  Specifically, we measured the time needed for one iteration of the algorithm to solve Poisson's equation on JuQueen.

  
    \item \textsc{JavaGC} is the \textit{Java Garbage Collector} of OpenJDK version 1.8.
  In our experiments, we consider configuration options varying, for example, garbage-collection policies or the heap space of the garbage collector, when performing the xalan benchmark from the DaCapo benchmark suite~\footnote{http://www.dacapobench.org/}.
  Overall, we measured the time spent during garbage collection for 193\,536 different configurations on an Intel Core i7-4770 @ 3.40 GHz with 4 CPU cores and 32 GB RAM (Ubuntu 16.04).

  \item \textsc{TriMesh} is a multigrid implementation operating on triangular grids to solve Poisson's equation.
  As configuration options, we consider four different smoothers and two different cycle types as well as variable numbers of pre-smoothing and post-smoothing steps (performed in one iteration of the multigrid algorithm).
  We also consider the interior angles of the triangles of the domain.
  Overall, we measured the time needed for one iteration on 239\,360 different configurations on a MacBook Pro with a Core i5 2.7GHz and 8GB RAM, running OS X 10.10 (Yosemite).
  
  \item \textsc{VP9} is an open video codec\footnote{https://www.webmproject.org/vp9/}.
  We consider configuration options that define, for example, the quality of the encoded video or the number of CPUs used in the conversion process.
  In our experiments, we consider version 1.5.0 to encode two seconds of the Big Buck Bunny video on 720p\footnote{https://media.xiph.org/}.
  Overall, we measured the time needed for the conversion of 216\,000 different configurations on an Intel Core i7-4770 @ 3.40 GHz with 4 CPU cores and 32 GB RAM.

\end{itemize}

Because of confounding factors, such as measurement bias, we measured each individual configuration multiple times until reaching a standard deviation of less than 10\,\% for its measured runtime.
To create the whole set of measurements, containing all learning sets and all remaining configurations of all systems for evaluation, we spent multiple years of computation time. 
That is, we provide the largest set of performance measurements in the domain of configurable software systems that we are aware of. 
All raw measurements are available at our supplementary Web site.

\setlength{\tabcolsep}{3pt}
\begin{table}
  \caption{Overview of the subject systems. \#$O_{Bin}$: Number of binary configuration options; \#$O_{Num}$: Number of numeric configuration options; \#C: Number of configurations measured in a brute force manner; \#Constr: Number of constraints among the configuration options.}
  \label{table:caseStudies}
  \centering
  \begin{tabular}{lrrrrl}
  \toprule
  System & \#$O_{\mathit{Bin}}$ & \#$O_{\mathit{Num}}$ & $\mathit{\#C}$ & $\mathit{\#Constr}$ & Performance metric\\
  \midrule
  \textsc{DUNE MGS}& 8 & 3 & 2\,304 & 13 & Time to solution\\
  \textsc{gemm} & 15 & 4 & 59\,592 & 11 & Runtime of complied program\\
  \textsc{HSMGP} & 11 & 3 & 3\,456 & 26 & Time for an iteration\\
  \textsc{JavaGC} & 5 & 6 & 193\,536 & 0 & Garbage collection time\\
  \textsc{TriMesh} & 9 & 4 & 239\,360 & 17 & Time for an iteration\\
  \textsc{VP9} & 15 & 5 & 216\,000 & 19 & Encryption time\\
  \bottomrule                                                  
  \end{tabular}
\end{table}
\setlength{\tabcolsep}{6pt}

\vspace{\baselineskip}

\subsection{Study Setup} \label{sec:studySetup}


To answer our research questions, we have performed more than \textit{15\,000 experiments} using each combination of a machine-learning technique with a binary sampling strategy and a numeric sampling strategy to predict the performance of all configurations of each subject system.
The setup is illustrated in Figure~\ref{fig:studySetup}.
In each experiment, we first use one binary and one numeric  sampling strategy to select a set of binary configurations $L_{\mathit{Bin}}$ and a set of numeric configurations $L_{\mathit{Num}}$. 
Then, we compute the cartesian product of the selected configurations to define the learning set $L = L_{\mathit{Bin}} \times L_{\mathit{Num}}$.

\begin{figure}
  \centering
    \includegraphics[width=0.65\textwidth]{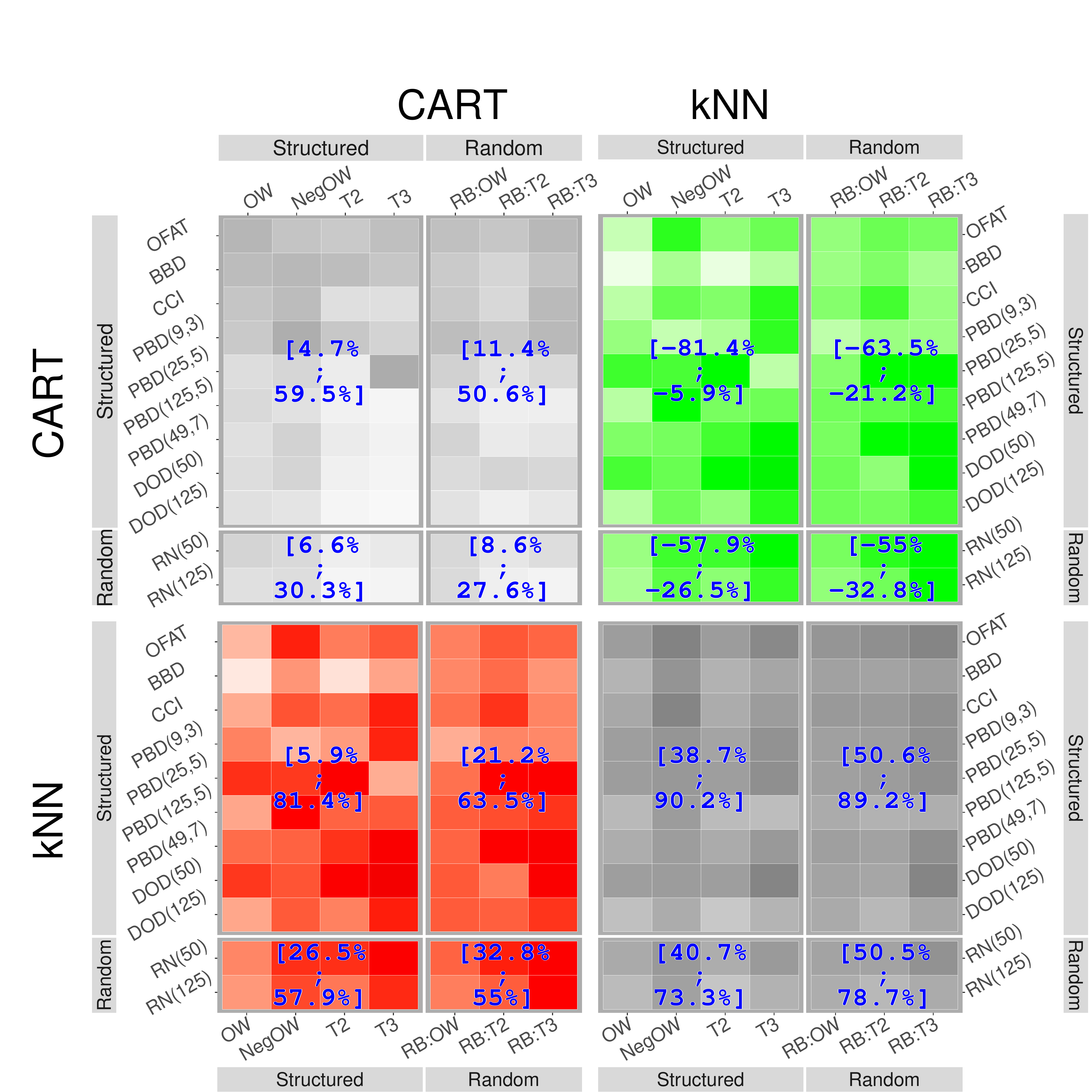}
  \caption{Excerpt from the comparison of the different machine-learning techniques (CART and kNN) in terms of the mean prediction error rate over all subject systems. On the diagonal, we plot the prediction error rate of the respective technique. For the plots not on the diagonal, we show the differences in the error rates, leading to green cells and negative numbers if the machine-learning technique in the row is more accurate than the one in the column, and red cells with positive numbers otherwise.}
  \label{fig:paretoExcerpt}
\end{figure}

We compare the mean error rates achieved when using two different machine-learning techniques or sampling strategies.
For illustration, in Figure~\ref{fig:paretoExcerpt}, we show a small excerpt of comparisons in the form of nested matrix plots, in which we compare the error rates achieved by CART and kNN (outer plot), when using different binary and numeric sampling strategies (inner plot). 
Each cell in one nested matrix plot is dedicated to one specific combination of a binary and a numeric sampling strategy. 

In all nested matrix plots, we distinguish between plots that are on the diagonal of the top-level matrix and plots that are not on the diagonal of the matrix.
In general, if a plot is not on the diagonal, we compare two different machine-learning techniques or sampling strategies respectively (column vs. row).
Plots on the diagonal present the error rates of the predictions when using a specific machine-learning technique or sampling strategy respectively without comparison.
For example, in the upper left plot in Figure~\ref{fig:paretoExcerpt}, we \textit{show} the error rates for CART and, in the plot in the upper right corner, we \textit{compare} the error rates achieved when using CART with the error rates of kNN.

When comparing the error rates of two machine-learning techniques (plots that are not on the diagonal), we compute the differences of their error rates leading to negative numbers if the machine-learning technique considered in the row is \textit{more} accurate than the machine-learning technique in the column, and to positive numbers otherwise (see for example, the plot in the upper right corner of Figure~\ref{fig:paretoExcerpt}, which shows that CART achieves smaller error rates than kNN).
We also color the cells based on the difference in the error rate of the predictions: 
if a cell is red, the machine-learning technique in the column achieves more accurate predictions than the one in the row, and green otherwise.
See, for example, the plot in the upper right corner of the matrix.
Based on the numbers and the colors in the plot, all being positive and red, we see that CART achieves smaller prediction errors than kNN; the differences are between 5.9\,\% and 81.4\,\%, depending on the used sampling strategies.
If the differences are small, the color of a cell fades into white. 
This is the case in the second row of the top-level plot in the upper right corner of Figure~\ref{fig:paretoExcerpt}, comparing the error rates of  CART and kNN, when using BBD for sampling numeric configuration options.
Looking at Figure~\ref{fig:paretoExcerpt}, we can easily conclude that CART outperforms kNN with respect to prediction accuracy independent to the sampling strategy. 

For plots on the diagonal, we present the error rates of the machine-learning technique considered in the row (and column).
Here, again, we encode the error rate when using a specific sampling strategy combination in the color of the dedicated cell.
For these plots, cells representing an experiment having a low error rate fade into white, whereas experiments with a large prediction error are represented as black cells.
For example, in the top-level plot in the upper right corner of Figure~\ref{fig:paretoExcerpt}, we see that DOD(50) in combination with T3 leads to more accurate predictions than PBD(25,5) in combination with T3.

In each nested (inner) matrix plot, we group random sampling strategies and strategies that purposefully select configurations based on coverage criteria for better comparison.
This can be seen in the upper left part of the top-level upper right plot of Figure~\ref{fig:paretoExcerpt}, where the experiments using CART outperform the experiments using kNN by an error rate of 81.4\,\% to 5.9\,\%, if structured-sampling strategies for both kinds of options are used.
In each of the plots, we have four groups of sampling strategy combinations, one where two structured sampling strategies are used (upper left part), one where structured sampling strategies for binary options and random sampling for numeric options are used (lower left corner), one where random sampling for binary options and structured sampling strategies for numeric options are used (upper right corner), and one where random sampling strategies are used for both binary and numeric options.

To test the significance of the differences in the error rates when using different machine-learning techniques or sampling strategies, we performed an one sided pairwise Wilcox test. For pairs with a significant difference in the error rate, we also calculate Cliff's delta describing the strength of the significance~\cite{Cliff1996}. In these tests, our null hypothesis is that the machine-learning technique on the column leads to smaller error rates as the machine-learning technique on the row. As a consequence, if our null hypothesis does not hold, we have negative values for Cliff's delta.

\begin{figure*}
  \centering
    \includegraphics[width=0.3\textwidth]{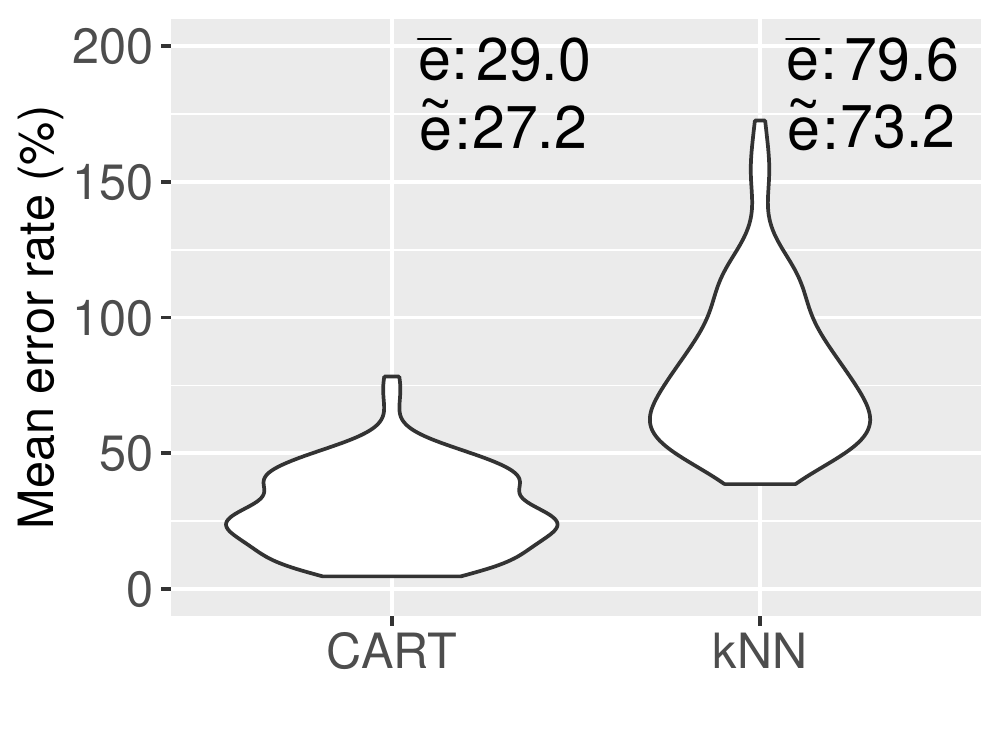}
  \caption{Excerpt from the robustness of machine-learning techniques ($\overline{e}$: mean value of the error; $\widetilde{e}$: median value of the error).}
  \label{fig:violin_learner_subset}
\end{figure*}

To compare the stability of the predictions when using the different machine-learning techniques and sampling strategies respectively, we use violin plots. 
Here, we consider the height as well as the thickness of the violins.
In Figure~\ref{fig:violin_learner_subset}, we provide a small excerpt of the violin plots describing the stability of the two machine-learning techniques kNN and CART.
Here, we see that CART is more stable compared to kNN, based on the thickness of the two violins, where the violin of CART is thicker at the thickest part compared to the violin of kNN.
Besides, the stability of CART can also be seen based on the height of the two violins, describing the distribution of the error rates.

\section{Results}

Guided by our research questions, we compare the different machine-learning techniques, binary-sampling strategies, numeric-sampling strategies, and combinations thereof in terms of their prediction accuracy.
In particular, we consider the mean prediction error of all configurations aggregated over all subject systems (RQ1.1 and RQ2.1).
Furthermore, we analyze the robustness of the observed prediction accuracy (RQ1.2 and RQ2.2).
Finally, we discuss outliers for which the influence of learning techniques or sampling strategies differ from the overall picture to obtain more insights how individual properties of software systems may change the picture.

\subsection{Comparison of Machine-Learning Techniques}

Next, we take a closer look at the influences of the individual machine-learning techniques on the error rate of the predictions.

\subsubsection{Mean Error Rate (RQ1.1)}

\begin{figure*}
  \centering
    \includegraphics[width=1.00\textwidth]{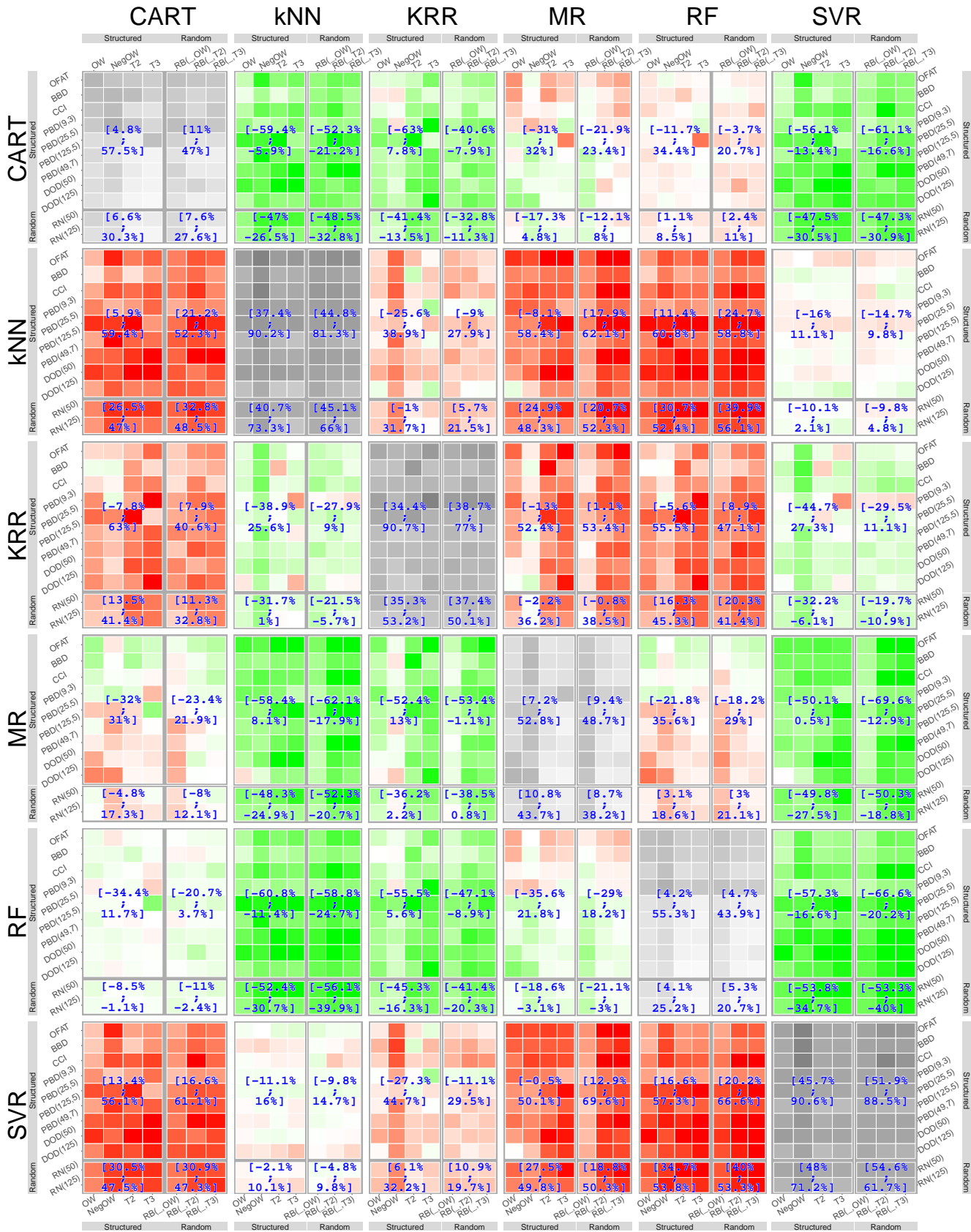}
  \caption{Comparison of machine-learning techniques by mean error rate when predicting all configurations of all subject systems.}
  \label{fig:pareto_Averaged_AveragedRandom_byLearner}
\end{figure*}

In Figure~\ref{fig:pareto_Averaged_AveragedRandom_byLearner}, we compare the machine-learning techniques regarding the mean prediction errors aggregated over all subject systems and grouped by the sampling strategy used.
As described in Section~\ref{sec:studySetup}, the results for one machine-learning technique are shown in one row and the respective column of the nested matrix plot.
In each of the nested matrix plots, we show the results dedicated to one binary sampling strategy in one column and the results when using one specific numeric sampling strategy in one row.

Looking at Figure~\ref{fig:pareto_Averaged_AveragedRandom_byLearner}, we notice two groups of machine-learning techniques, one group predicts performance with a low error (green), and one group yields large prediction errors (red).
The first group contains CART, RF, and MR, which outperform the other techniques with respect to prediction accuracy for almost all binary and numeric-sampling strategies.
These three learning techniques are able to predict the performance of all configurations with a mean prediction error of between 4.1\,\% and 55.3\,\%, depending on the selected sampling strategy (see the plots on the diagonal of Figure~\ref{fig:pareto_Averaged_AveragedRandom_byLearner} for CART, RF, and MR).
When comparing the error rates of these three techniques, we see that none of them is superior to the others, across all experiments.
However, based on the differences of error rates (and the colors of the cells), we see that RF leads to slightly better predictions compared to the other two learning techniques, on average.

The second group consists of KRR, kNN, and SVR, yielding larger error rates than the first group, independently of the sampling strategy.
When using KRR, we observe error rates of more than 34.4\,\%, on average, while the other two lead to even worse predictions: 37.4\,\% error for kNN and 45.7\,\% for SVR.

In Table~\ref{table:sigLearner}, we present the p-values when comparing the mean error rates of two machine-learning techniques. 
For significant differences, we also calculated Cliff's delta describing the strength of the difference.
Overall, we see that CART, RF, and MR lead to significantly smaller error rates than kNN, KRR, and SVR.
Based on Cliff's delta, we see that all of these differences have a large effect size ($< -0.474$) . 
There is also a significant difference between RF and CART or MR.
Here, based on the effect sizes, we see that although there are significantly difference in the error rate, the difference only has a small effect size.
For the group of machine-learning techniques with a large error rate, we see that there is a significant difference with a large effect size between KRR and SVR, one significant difference with a medium effect size between KRR and kNN, and a significant difference with a small effect size between kNN and SVR.



\setlength{\tabcolsep}{3pt}
\begin{table}
  \caption{
 p-values from a one sided pairwise Wilcox test, where we tested pair-wisely whether the machine-learning technique on the row leads to significant smaller error rates compared to the machine-learning technique in the column. For significant differences, we present the strength of the differences in terms of Cliff's delta; we highlighted the cells with a strong effect size \vertexStrong, a medium effect size \vertexMedium, and a small effect size \vertexSmall.}
  \label{table:sigLearner}
  \centering
  \begin{tabular}{p{0.14\textwidth-2\tabcolsep}p{0.14\textwidth-2\tabcolsep}p{0.14\textwidth-2\tabcolsep}p{0.14\textwidth-2\tabcolsep}p{0.14\textwidth-2\tabcolsep}p{0.14\textwidth-2\tabcolsep}p{0.14\textwidth-2\tabcolsep}}
  \toprule
  &CART & kNN & KRR & MR & RF & SVR\\ 
      \midrule 
CART & --- & \cellcolor{sigLarge} 2.33e-26\newline(-0.946) & \cellcolor{sigLarge} 3.68e-21\newline(-0.836) & 0.90580 & 1.00 & \cellcolor{sigLarge} 3.15e-27\newline(-0.974)\\ 
kNN & 1.00000 & --- & 1.00e+00 & 1.00000 & 1.00 & \cellcolor{sigSmall} 3.27e-02\newline(-0.206)\\ 
KRR & 1.00000 & \cellcolor{sigMedium} 1.02e-04\newline(-0.340) & --- & 1.00000 & 1.00 & \cellcolor{sigLarge} 8.92e-08\newline(-0.477)\\ 
MR & 0.99334 & \cellcolor{sigLarge} 1.27e-20\newline(-0.830) & \cellcolor{sigLarge} 9.60e-16\newline(-0.728) & --- & 1.00 & \cellcolor{sigLarge} 5.41e-22\newline(-0.870)\\ 
RF & \cellcolor{sigSmall} 0.00413\newline(-0.278) & \cellcolor{sigLarge} 2.09e-27\newline(-0.971) & \cellcolor{sigLarge} 3.47e-24\newline(-0.913) & \cellcolor{sigSmall} 0.00584\newline(-0.263) & --- & \cellcolor{sigLarge} 2.09e-27\newline(-0.984)\\ 
SVR & 1.00000 & 1.00e+00 & 1.00e+00 & 1.00000 & 1.00 & ---\\ 

  \bottomrule                                                  
  \end{tabular}
\end{table}
\setlength{\tabcolsep}{6pt}

\subsubsection{Stability (RQ1.2)} \label{sec:result:stabML}

\begin{figure*}
  \centering
    \includegraphics[width=\textwidth]{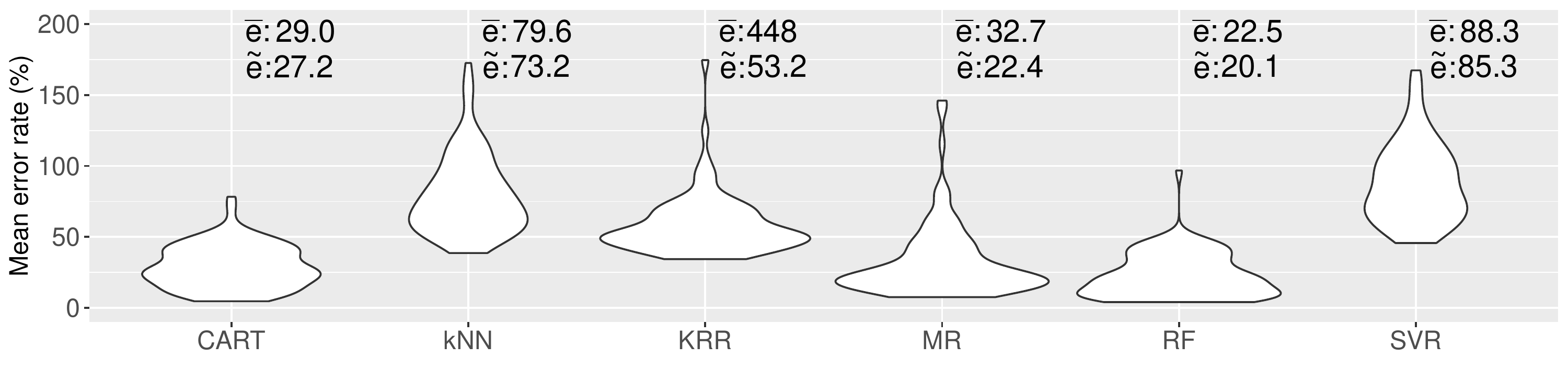}
  \caption{Robustness of machine-learning techniques ($\overline{e}$: mean value of the error; $\widetilde{e}$: median value of the error). We do not show the outliers from KRR affecting the mean error rate of the predictions but consider them in the computation of the mean error rate.}
  \label{fig:violin_learner}
\end{figure*}

To identify the machine-learning techniques that produce the most stable results in terms of low variation of prediction errors across subject systems and sampling strategies, we present the respective distributions of error rates in Figure~\ref{fig:violin_learner} (we provide the minimal and maximal error rates in the supplementary Web site).
For all distributions, we include the mean value of the error ($\overline{e}$) and the median value ($\widetilde{e}$).
In general, when comparing stability, we have to consider the distribution of the thickness (i.e., density) as well as the height (i.e., variation) of the violin plots and the minimal and maximal error rates.

Based on Figure~\ref{fig:violin_learner}, we notice three groups of machine-learning techniques. 
First, we observe that RF and CART lead to the most stable results (both of their violins have a smaller height compared to the violins of the other techniques).
In the second group, we see that KRR and MR are stable for a large number of experiments, but also lead to large error rates for other runs.
Last, we see that kNN and SVR are the least stable techniques.
When, considering the minimal and maximal error rates, we see that RF and CART are more stable than any other learning techniques.
For the other learning techniques, we see that all of them strongly depend on the learning set of configurations.


\subsubsection{Outliers} \label{sec:result:ML:Outliers}

When analyzing the results per subject system, we obtain a more diverse picture.
We provide all figures for these comparisons at our supplementary Web site\footnote{https://www.infosun.fim.uni-passau.de/se/projects/splconqueror/expDesign.php}.
In what follows, we discuss the most notable observations.

For \textsc{Dune MGS}, we observe that all techniques are comparably accurate; as all of them reach a mean prediction error of less than 11\,\% if suitable sampling strategies are used.
This includes even techniques that we found to be very inaccurate in Section~\ref{sec:result:stabML} (kNN, SVR, and KRR).
One reason for that might be the small variation of the performance of \textsc{Dune MGS} and the relative simple influences of configuration options and interactions. 

For \textsc{gemm}, we see that, in some experiments, kNN achieves more accurate predictions, when predicting all configurations, compared to CART, RF, and MR, which is in contrast to the overall picture. 
In these experiments, the binary sampling strategies OW, NegOW, and RB(\_,OW) are used and the largest error rates of all experiments were achieved. 
One reason for the large error rates when using one of these three sampling strategies is the large influence of the binary configuration options on performance.

\subsubsection{Summary}

Overall, we found that CART, RF, and MR are able to actually predict the performance of all configurations, while kNN, KRR, and SVR lead to larger error rates.
We also see that the error rates of all learning techniques strongly depend on the choice of the sampling strategy.
Thus, even a learning technique that produces accurate predictions, in general, can have a high error rate if a unsuitable sampling strategy is used.

\begin{figure*}
  \centering
    \includegraphics[width=\textwidth]{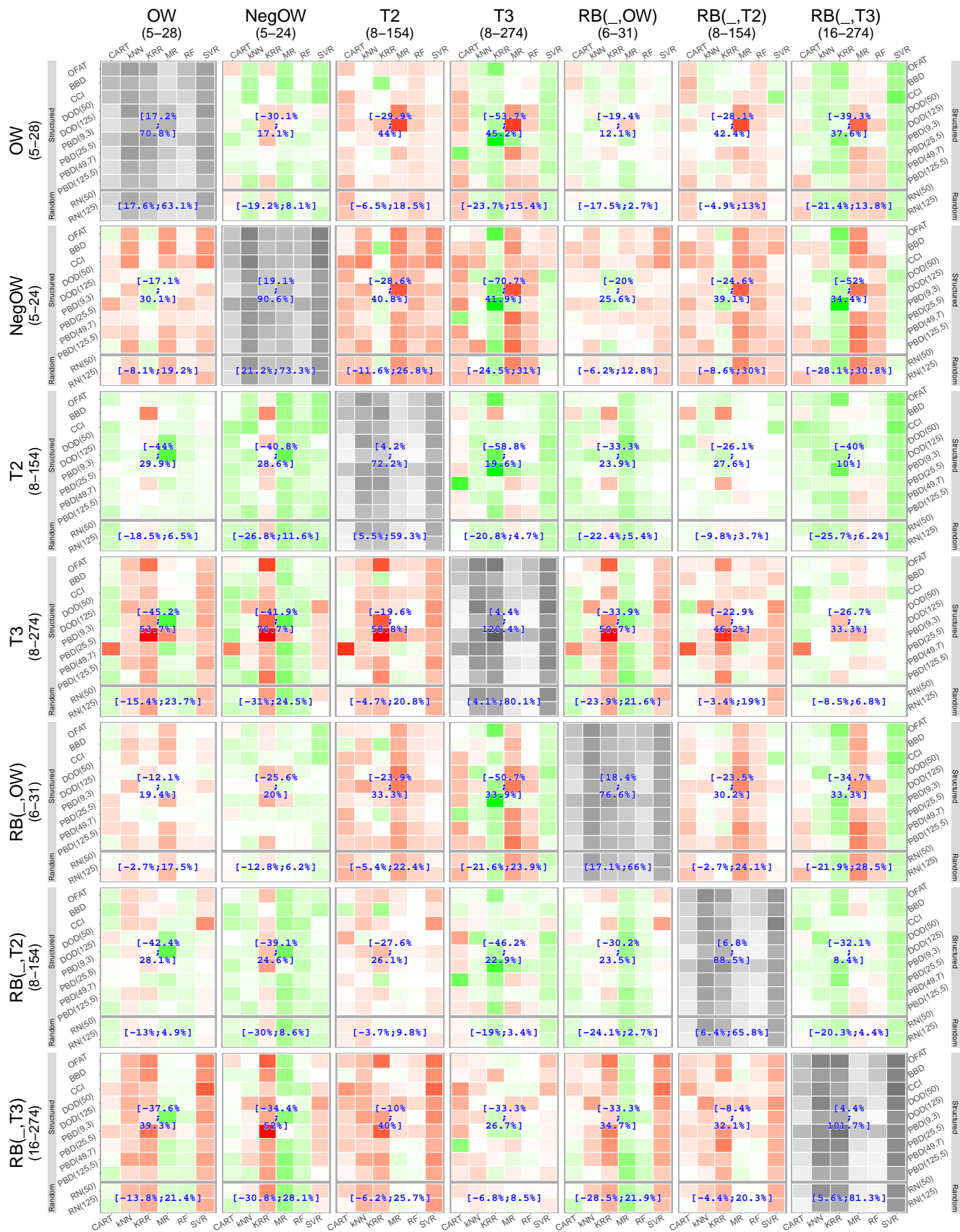}
  \caption{Comparison of binary-sampling strategies by mean error rate when predicting all configurations of all subject systems.}
  \label{fig:pareto_Averaged_AveragedRandom_byBinSamp}
\end{figure*}


\subsection{Comparison of Binary-Sampling Strategies}\label{sec:comBin}

Next, we consider the influence of the binary-sampling strategies on prediction accuracy.

\subsubsection{Mean Error Rate (RQ2.1)}

In Figure~\ref{fig:pareto_Averaged_AveragedRandom_byBinSamp}, we compare the binary-sampling strategies regarding the mean error rate when predicting all configurations of the subject systems.
In contrast to Figure~\ref{fig:pareto_Averaged_AveragedRandom_byLearner}, we now show the binary-sampling strategies at the top level of the nested matrix plot.
We also show the number of selected configurations in the header of the nested matrix plot.
For example, when using the T2 sampling strategy, we select 8 to 154 different configurations depending on the respective subject system.
In Table~\ref{table:sigBinary}, we show the p-values and effect sizes, when comparing the error rates achieved using different binary sampling strategies.

An important aspect we have to take into account in this comparison is the number of selected configurations determined by the individual sampling strategies.
Clearly, more measurements should yield more accurate predictions.
Hence, we compare three groups of binary-sampling strategies: few measurements $\mathcal{B}_{\mathit{small}}$ (OW, NegOW, and RB(\_,OW)), medium number of measurements $\mathcal{B}_{\mathit{med}}$ (T2 and RB(\_,T2)), and many measurements $\mathcal{B}_{\mathit{large}}$ (T3 and RB(\_,T3)).

For $\mathcal{B}_{\mathit{small}}$, we observe that OW leads to the smallest error rates for most experiments. 
Looking at Table~\ref{table:sigBinary}, we see that difference between OW and NegOW is significant with a small effect size.
For $\mathcal{B}_{\mathit{med}}$, T2 leads to more accurate predictions compared to RB(\_,T2), for almost all experiments. 
However, although there is a difference in error rates, it is not significant (Table~\ref{table:sigBinary}).

For $\mathcal{B}_{\mathit{large}}$, none of the strategies outperforms any other, in general, which is also stated in Table~\ref{table:sigBinary}.

Overall, we see that using T2 leads mostly to predictions with the smallest error rate: between 4.2\,\% and 72.2\,\%, depending on the applied numeric sampling strategy and the used machine-learning technique.
However, we also see that none of the binary sampling strategies is superior, in general (compare the distribution of red and green in Figure~\ref{fig:pareto_Averaged_AveragedRandom_byBinSamp}).
For each binary sampling strategy, we see that there is, at least, one experiment, where it leads to more accurate predictions than any other binary sampling strategy.
This even holds for NegOW and RB(\_,OW) leading to large error rates for a large number of experiments, but to the most accurate predictions for other combinations of machine-learning technique and numeric sampling strategy (see, for example, the results for OFAT and CART).

\setlength{\tabcolsep}{3pt}
\begin{table}
  \caption{p-values from a one sided pairwise Wilcox test, where we tested pair-wisely whether the binary sampling strategy on the row leads to significant smaller error rates compared to the binary sampling strategy in the column. For significant differences, we present the strength of the differences in terms of Cliff's delta; we highlight strong effect sizes \vertexStrong, medium effect sizes \vertexMedium, and small effect sizes \vertexSmall.}
  \label{table:sigBinary}
  \centering
  \begin{tabular}{p{0.11\textwidth-2\tabcolsep}p{0.11\textwidth-2\tabcolsep}p{0.11\textwidth-2\tabcolsep}p{0.11\textwidth-2\tabcolsep}p{0.11\textwidth-2\tabcolsep}p{0.11\textwidth-2\tabcolsep}p{0.11\textwidth-2\tabcolsep}p{0.11\textwidth-2\tabcolsep}p{0.11\textwidth-2\tabcolsep}}
  \toprule
  &OW&NegOW&T2&T3&RB(\_,OW)&RB(\_,T2)&RB(\_,T3)\\ 
      \midrule
OW&---&\cellcolor{sigSmall} 0.01969\newline(-0.270)&1.000&1.000&0.19919&1.000&1.000\\ 
NegOW&1.000&---&1.000&1.000&0.99994&1.000&1.000\\ 
T2&0.083&\cellcolor{sigMedium} 0.00125\newline(-0.387)&---&0.558&\cellcolor{sigMedium} 0.00125\newline(-0.371)&0.290&0.112\\ 
T3&0.777&0.07325&0.991&---&0.09794&0.777&0.630\\ 
RB(\_,OW)&0.991&0.50035&1.000&1.000&---&1.000&1.000\\ 
RB(\_,T2)&0.389&\cellcolor{sigSmall} 0.01859\newline(-0.281)&1.000&1.000&\cellcolor{sigSmall} 0.01978\newline(-0.262)&---&1.000\\ 
RB(\_,T3)&0.630&\cellcolor{sigSmall} 0.04443\newline(-0.209)&1.000&1.000&0.06638&0.777&---\\ 
  \bottomrule                                                  
  \end{tabular}
\end{table}
\setlength{\tabcolsep}{6pt}

%
%
%

\subsubsection{Stability (RQ2.2)}
\begin{figure*}
  \centering
    \includegraphics[width=\textwidth]{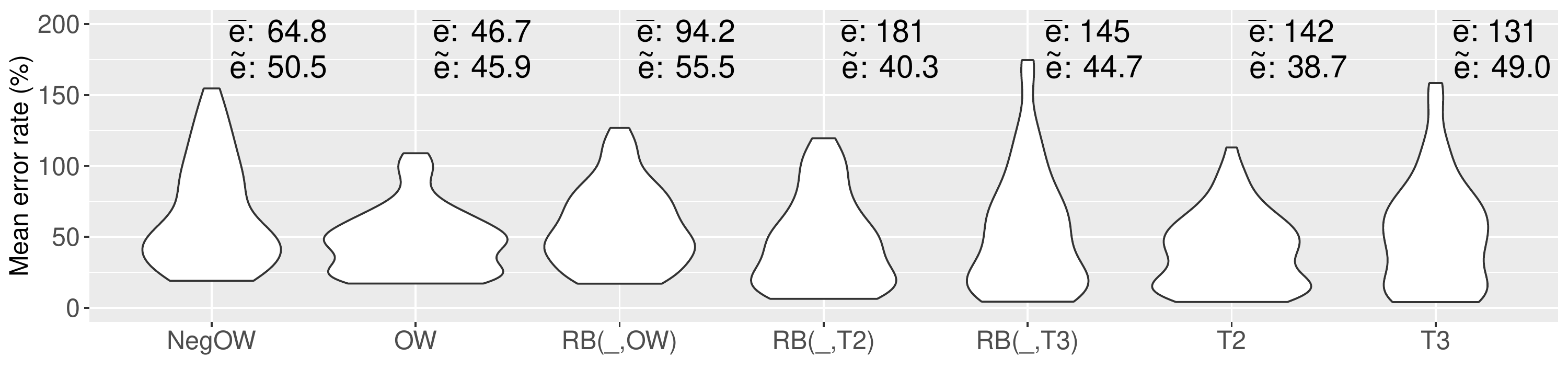}
  \caption{Robustness of the binary sampling strategies ($\overline{e}$: mean value of the error; $\widetilde{e}$: median value of the error).}
  \label{fig:violin_bin}
\end{figure*}

In Figure~\ref{fig:violin_bin}, we present the distribution of the error rates for the different binary-sampling strategies and the minimal and maximal values on our supplementary Web site.
For the distribution of the error rates, we observe that OW leads to the most stable results (its violin plot has a lower height and is also thicker at its thickest part and they also lead to a smaller distribution of the error rates).
Furthermore, we see that RB(\_,T3) and T3 are the least stable sampling strategies.
This is a surprising result as more measurements should result in more stable predictions. 
We investigate this observation further in Section~\ref{sec:discuss:r21}.


\subsubsection{Outliers}\label{sec:result:bin:Out}

Let us now take a closer look at individual systems, for which the results differ from the results obtained from an aggregation.

For \textsc{HSMGP}, we observe that the choice of the binary-sampling strategy has only a minor influence on the mean error rate compared to their influence when considering all subject systems.
However, we see that using RB(\_,OW) leads to higher error rates compared to one of the other binary sampling strategies. 
Here, all binary sampling strategies, except for RB(\_,OW), can lead to prediction errors smaller than 1.5\,\%, while the smallest prediction error of RB(\_,OW) is about 8.6\,\%.
After having a deeper look at the learning sets defined by the different seeds we use in RB(\_,OW), we saw that one configuration option is not considered in half of the learning sets.
As a consequence, this option, although having only a small influence on performance, is not considered by the machine-learning techniques in half of the experiments when using RB(\_,OW), which explains the larger error rate when using this sampling strategy. 

For \textsc{gemm}, we obtain a different picture.
Here, OW leads to larger errors compared to NegOW, for all combinations of machine-learning techniques and numeric-sampling strategies, except for KRR when applying sampling strategies of $\mathcal{B}_{\mathit{med}}$ and $\mathcal{B}_{\mathit{large}}$.
This is in stark contrast to the overall picture, where NegOW leads to the largest error rates of all sampling strategies.
A closer look at the performance-influence models that we learned for \textsc{gemm}, reveals that there are some interactions among binary configuration options with a considerable influence on the performance.
Although these interactions are only of order 2 and 3, they contain almost all configuration options that influence performance.

\subsubsection{Summary}

Our experiments show that the benefits of using structured sampling strategies, as compared to a random-based selection of the same number of configurations, decreases with an increasing number of selected configurations.
Furthermore, we see that NegOW and RB(\_,OW) lead to the largest error rates in most experiments.
However, this general observation does not hold for all individual subject systems; there are cases where NegOW leads to more accurate predictions than OW sampling, for example.


\subsection{Comparison of Numeric-Sampling Strategies}\label{sec:comNum}

Next, we evaluate the influence of numeric-sampling strategies on the accuracy of performance predictions.

\subsubsection{Mean Error Rate (RQ2.1)}

\begin{figure*}
  \centering
    \includegraphics[width=\textwidth]{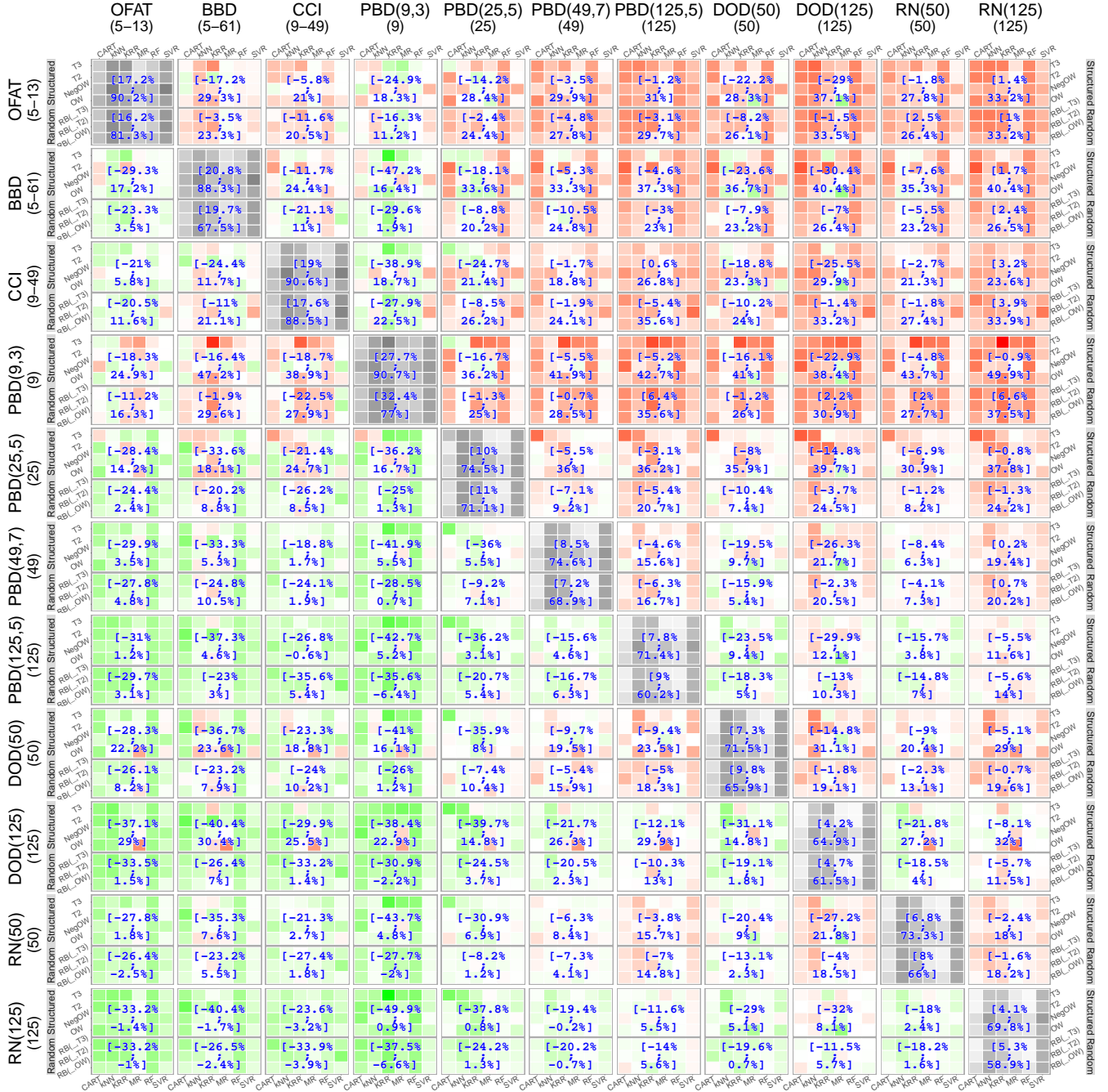}
  \caption{Comparison of the mean error rates of different numeric-sampling strategies of all subject systems.}
  \label{fig:pareto_Averaged_AveragedRandom_byNumSamp}
\end{figure*}

\tiny
\setlength{\tabcolsep}{3pt}
\begin{table}
  \caption{p-values from a one sided pairwise Wilcox test, where we tested pair-wisely whether the numeric sampling strategy on the row leads to significant smaller error rates compared to the numeric sampling strategy in the column. For significant differences, we present the strength of the differences in terms of Cliff's delta; we highlight strong effect sizes \vertexStrong, medium effect sizes \vertexMedium, and small effect sizes \vertexSmall.}
  \label{table:sigNumeric}
  \centering
  \begin{tabular}{p{0.08\textwidth-2\tabcolsep}p{0.08\textwidth-2\tabcolsep}p{0.08\textwidth-2\tabcolsep}p{0.08\textwidth-2\tabcolsep}p{0.08\textwidth-2\tabcolsep}p{0.08\textwidth-2\tabcolsep}p{0.08\textwidth-2\tabcolsep}p{0.08\textwidth-2\tabcolsep}p{0.08\textwidth-2\tabcolsep}p{0.08\textwidth-2\tabcolsep}p{0.08\textwidth-2\tabcolsep}p{0.08\textwidth-2\tabcolsep}}
  \toprule
  &OFAT&BBD&CCI&PBD(9,3)&PBD(25,5)&PBD(125,5)&PBD(49,7)&DOD(50)&DOD(125)&RN(50)&RN(125)\\
      \midrule
OFAT&---&1.00000&0.99999&1.89e-01&1.00000&1.000&1.0000&1.0000&1.000&1.000&1.00\\ 
BBD&0.573287&---&0.99999&7.98e-02&0.99999&1.000&1.0000&1.0000&1.000&1.000&1.00\\ 
CCI&0.489908&0.94301&---&\cellcolor{sigSmall} 4.49e-02\newline(-0.287)&1.00000&1.000&1.0000&1.0000&1.000&1.000&1.00\\ 
PBD(9,3)&0.999994&1.00000&0.99999&---&0.99999&1.000&1.0000&1.0000&1.000&1.000&1.00\\ 
PBD(25,5)&0.286132&0.67113&0.75776&\cellcolor{sigSmall} 2.47e-02\newline(-0.321)&---&1.000&1.0000&0.8613&1.000&1.000&1.00\\ 
PBD(125,5)&\cellcolor{sigMedium} 0.004233\newline(-0.406)&\cellcolor{sigMedium} 0.01110\newline(-0.366)&\cellcolor{sigMedium} 0.00774\newline(-0.381)&\cellcolor{sigLarge} 2.60e-04\newline(-0.537)&\cellcolor{sigSmall} 0.03272\newline(-0.305)&---&0.2394&0.1245&1.000&0.292&1.00\\ 
PBD(49,7)&0.056443&0.18905&0.12454&\cellcolor{sigMedium} 3.14e-03\newline(-0.427)&0.25997&1.000&---&0.4437&1.000&1.000&1.00\\ 
DOD(50)&0.380009&0.48991&0.67821&6.32e-02&1.00000&1.000&1.0000&---&1.000&1.000&1.00\\ 
DOD(125)&\cellcolor{sigMedium} 0.003892\newline(-0.408)&\cellcolor{sigMedium} 0.01492\newline(-0.349)&\cellcolor{sigMedium} 0.00954\newline(-0.363)&\cellcolor{sigLarge} 2.60e-04\newline(-0.524)&\cellcolor{sigSmall} 0.02310\newline(-0.329)&0.822&0.1891&0.0842&---&0.239&1.00\\ 
RN(50)&\cellcolor{sigSmall} 0.022760\newline(-0.319)&0.09958&0.09342&\cellcolor{sigLarge} 8.35e-04\newline(-0.474)&0.12454&1.000&0.7578&0.2861&1.000&---&1.00\\ 
RN(125)&\cellcolor{sigLarge} 0.000835\newline(-0.477)&\cellcolor{sigMedium} 0.00239\newline(-0.435)&\cellcolor{sigMedium} 0.00111\newline(-0.452)&\cellcolor{sigLarge} 2.37e-05\newline(-0.596)&\cellcolor{sigMedium} 0.00954\newline(-0.359)&0.573&0.0874&\cellcolor{sigSmall} 0.0393\newline(-0.299)&0.722&0.127&---\\ 

  \bottomrule                                                  
  \end{tabular}
\end{table}
\setlength{\tabcolsep}{6pt}
\normalsize

In Figure~\ref{fig:pareto_Averaged_AveragedRandom_byNumSamp}, we show the numeric-sampling strategies at the top level of the nested matrix plot.
We also present the number of selected configurations in the header of the matrix.
Again, the selection of the sampling strategy has a strong influence on the number of selected configurations.
However, in contrast to the binary-sampling strategies, we cannot easily partition the sampling strategies into groups, so we will discuss them separately.
We also present p-values and the related effect size when comparing the error rates achieved using different numeric sampling strategies in Table~\ref{table:sigNumeric}.

Overall, we see that OFAT, BBD, CCI, and PBD(9,3) lead to predictions with a high error rate caused by the small number of configurations selected by these strategies.
For example, when using PBD(9,3), only 9 different configurations are selected over the numeric configuration space, which is only $1/5$ of the configurations selected by other strategies.
When comparing OFAT with PBD(9,3) (both of them select only a small number of configurations), OFAT is more suitable in combination with KRR, MR, and RF, whereas PDB(9,3) yields more accurate predictions when combined with SVR.
In Table~\ref{table:sigNumeric}, we see that there are significant differences mostly between sampling strategies that select different numbers of configurations, such as PBD(125,5) and PBD(9,3).

In general, we observe that DOD(125), PBD(125,5), and RN(125) lead to the most accurate predictions for most of the learning techniques and binary-sampling strategies.
This is also apparent in Table~\ref{table:sigNumeric}: there is no significant difference in the error rates when using one of these three sampling strategies.
However, there is no clear winner, since each strategy outperforms another one for certain combinations of binary-sampling strategies and learning techniques.
This even holds for sampling strategies selecting only a small number of configurations, such as PBD(9,3) or OFAT.
When comparing the numeric sampling strategies selecting a small number of configurations with the sampling strategies selecting a large number of configurations, we see that there are significant differences with large or medium effect sizes.

\subsubsection{Stability (RQ2.2)}

\begin{figure*}
  \centering
    \includegraphics[width=\textwidth]{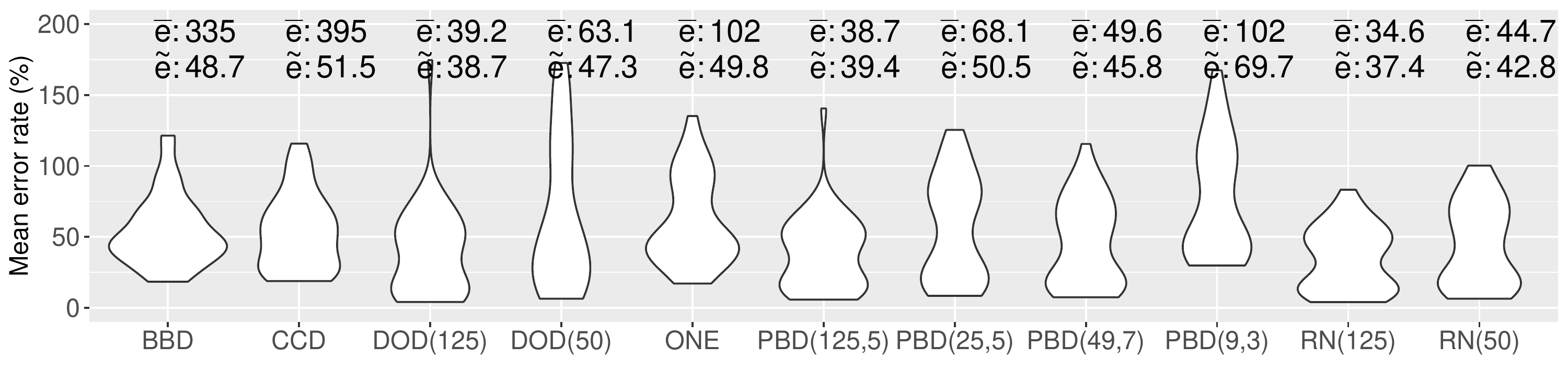}
  \caption{Robustness of numeric sampling strategies ($\overline{e}$: mean value of the error; $\widetilde{e}$: median value of the error).}
  \label{fig:violin_num}
\end{figure*}

In Figure~\ref{fig:violin_num}, we present the distribution of the error rates when using different numeric-sampling strategies.
DOD(50) and PBD(9,3) lead to the least stable results, whereas DOD(125), PBD(125,5), RN(125), BBD, and CCD provide stable results across the learning techniques. 
However, for both CCD and BBD we see some experiments where both of them lead to large error rates having a considerable influence on the stability of the experiments.

\subsubsection{Outliers}\label{sec:result:num:Out}

We found only one outlier in our data.
In most experiments on \textsc{gemm}, the selection of the numeric-sampling strategy has almost no influence on the mean error rates of the predictions.
Only for KRR, the selection of a specific numeric-sampling strategy has a large influence on the error.

\subsubsection{Summary}

Overall, for the numeric-sampling strategies, we see no benefits in using a structured-sampling strategy compared to randomly selecting configurations.
Moreover, we found the expected relationship between numbers of measurements and prediction accuracy: 
More measurements results in more accurate predictions no matter what learning technique is used.

\section{Discussion}

\subsection{Comparison of Machine-Learning Techniques}

\subsubsection{Mean Error Rate (\textsc{RQ1.1})}

Overall, none of the machine-learning techniques dominates other learning techniques in terms of a lower mean error rate, for all sampling strategy combinations on all subject systems. 
However, our results show that RF, CART, and MR leads to significant smaller error rates with a strong effect size compared to kNN, KRR, and SVR, independent of the sampling strategy.
To explain the larger error rates of kNN, KRR, and SVR, we have taken a closer look at their predictions and their internal mechanics.

One reason for the high error rates produced by kNN is the distance metric that should reflect the influences of the configuration options.
However, since these influences are unknown beforehand, standard metrics, such as the Euclidean distance have to be used, with which, all options are considered as equally important.
Furthermore, interactions among configuration options are not considered by such metrics and thus they are not considered in the prediction process.
That is, the non-linearity of the performance behavior introduced by interactions renders a kNN approach inaccurate.
The high error rates of KRR and SVR, which use transformations of the configuration space to learn linear models, can be explained by influences among options that are not considered in the transformation process, because they do not exist in the learning set.
As a consequence, we suggest that using either RF, CART, or MR because they lead to smaller error rates compared to kNN, SVR, and KRR.
They outperform the other techniques for different sampling strategy combinations on different subject systems. 
But, when comparing the strength of the differences using a Wilcox test and Cliff's delta, we see that there is either no significance difference (between CART and MR) or a significant difference with a small effect size only (between RF and CART and RF and MR).

A further finding is that it depends on the subject system and the used sampling strategy which specific learning technique performs best.
To this end, we analyzed the variation of the performance of the subject systems.\footnote{We define the performance variation of a system as $\frac{\mathit{max}-\mathit{min}}{\mathit{min}}$, where $\mathit{min}$ is the measured performance of the fastest configuration and $\mathit{max}$ the measured performance of the slowest configuration respectively.}
Here, we see that \textsc{HSMGP} and \textsc{gemm} exhibit the smallest performance variation among the configurations of our subject systems ($2.72$ and $6.44$ respectively).
On our supplementary Web site, we present the performance variation of all subject systems.
This might explain some outlier observations that we made for \textsc{gemm} and \textsc{Dune}, such as the small influence of the machine-learning techniques on the prediction accuracy or the small influence of the binary sampling strategies.

Beside performance variation, there are also other case specific characteristics that have an influence on the error rate of the different machine-learning techniques.
For example, when comparing the error rates of the different machine-learning techniques for \textsc{HSMGP} and \textsc{Dune}, which have a performance variation of $2.72$ and $12.13$ respectively, we see that, for most of the experiments, higher error rates were achieved for \textsc{HSMGP} although it has a smaller performance variation.
Having a closer look at the performance models that were learned, we see that the performance models of \textsc{TriMesh} consider only pairwise interactions, while the performance model of \textsc{VP9} also considers higher-order interactions (up to 5 interacting configuration options). 
Furthermore, while numeric configuration options have only a linear influence in \textsc{TriMesh}, there are higher-order polynomial functions (up to an order of 4) in the model of \textsc{VP9}.
When having such non-linear influences, kNN and SVR have large problems, whereas all other learning techniques are able to perform accurate predictions if a suitable learning set is selected.

So overall, for systems with a small performance variation and low interaction degree, the influence of the machine-learning technique becomes less relevant, which can, for example, be seen when inspecting the error rates of \textsc{Dune MGS}. 

\Recommendation{We recommend using RF, MR, or CART at first.
 If these techniques lead to large error rates, we suggest switching using another sampling strategy.}

\subsubsection{Stability (\textsc{RQ1.2})}

When comparing the stability of the mean error rates of the machine-learning techniques, we see that CART and RF are more stable than the other machine-learning techniques. 
This can be derived from the form of the violins and the minimal and maximal mean error rate of the learning techniques (provided on the supplementary Web site).
The main difference between CART and RF, compared to MR, KRR, and SVR, is that CART and RF do not extrapolate the influences of configuration options on regions that are not covered by the learning set.
This means that the predicted performance values of CART and RF are always within the value domain of the learning set. 
Overall, while extrapolation is often beneficial, it also comes with drawbacks, especially if the learning set does not represent the whole configuration space appropriately.
As a consequence, inappropriate functions are used in the transformation process.
Thus, extrapolation can lead to large prediction errors if a non-suitable learning set is selected, while it leads to good predictions if the learning set is selected appropriately.
For stability this means that the predictions of learning techniques that perform an extrapolation are more affected by the choice of the learning set.

\Recommendation{To achieve stable results independent to the learning set, we recommend using RF or CART, because they provide the most stable results across all systems.}

\subsection{Comparison of Sampling Strategies}

\subsubsection{Mean Error Rate (\textsc{RQ2.1})}\label{sec:discuss:r21}

\begin{figure}
  \centering
    \includegraphics[width=1\textwidth]{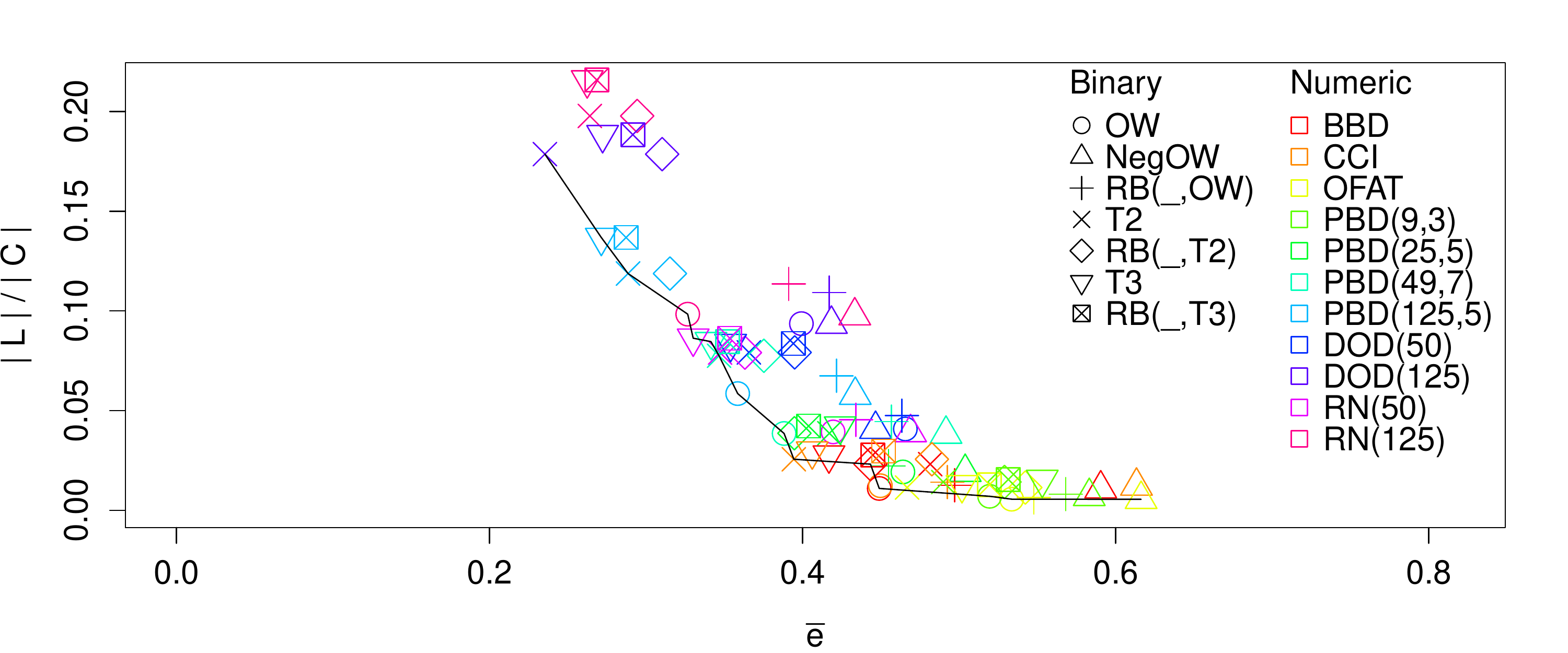}
  \caption{Relative size of the learning set ($|L| / |C|$) and mean error rate ($\overline{e}$) for all binary and numeric sampling strategy combinations aggregated over all subject systems. The combinations of the Pareto frontier are (ordered ascending by mean prediction error): T2-DOD(125), T3-PBD(125,5), T2-PBD(125,), OW-RN(125), T3-RN(50), T3-PBD(49,7), T2-PBD(49,7), OW-PBD(125,5), OW-PBD(49,7), T2-CCI, RB(\_,T2)-BBD, OW-BBD, OW-PBD(9,3), OW-OFAT, and NegOW-OFAT.}
  \label{fig:paretoFront_binary_average_rankingForBestConfig_FALSE}
\end{figure}

As shown in Figures~\ref{fig:pareto_Averaged_AveragedRandom_byBinSamp} and \ref{fig:pareto_Averaged_AveragedRandom_byNumSamp}, the choice of the sampling strategy has a strong influence on the mean error rate. 
To learn about the effectiveness of specific binary and numeric sampling-strategy combinations, we have to consider the tradeoff between the number of selected configurations and the error rate of the predictions. 
To this end, we show the Pareto-optimal set of sampling-strategy combinations in Figure~\ref{fig:paretoFront_binary_average_rankingForBestConfig_FALSE}, considering the mean error rate over all subject systems.
Similar plots for each individual subject system can be found at our supplementary Web site.

When comparing binary-sampling strategies without considering the size of the learning set, we see that NegOW leads to the largest error rates in many cases, whereas T2 leads to the smallest error rates.
The good results of T2 arise from the existence of performance-relevant pairwise interactions among the binary-configuration options of the systems.
Higher-order interactions (e.g., interactions among three options, which are covered by T3) have only a small influence on performance, if at all.
The nature of NegOW is to select configurations containing as many interactions as possible. 
This strategy, however, hinders a learner from assigning performance influences to individual options and interactions as they overlap.

When also taking the sample size into account, we see that structured sampling strategies for binary options lead to more accurate results than random strategies.
To make this point clear let us have a closer look at the sampling-strategy combinations on the Pareto frontier when averaging the error rate and the relative number of selected configurations over all subject systems. 
On our supplementary Web site, we provide detailed tables on which sampling-strategy combination is on the Pareto frontier of which subject system.
We see that only 1 out of 15 sampling strategy combinations on the Pareto front uses random sampling for binary configuration options, which is in stark contrast to published research papers mainly relying on random sampling~\cite{DBLP:conf/sigsoft/OhBMS17, DBLP:journals/corr/abs-1801-02175, Guo2013}.
We also see that each of the structured sampling strategies outperforms the other sampling strategies for, at least, one learning set size.
This even holds for NegOW, which leads to a larger error rate compared to the other sampling strategies, in general.
However, since it selects a smaller number of configurations compared to OW\footnote{This is the case for \textsc{gemm} and \textsc{VP9} and comes from the structure of the variability model in combination with the selection strategy that is performed by OW and NegOW.}, it is also on the Pareto frontier.
Having a closer look, though, we can see that this is only due a small decrease in selected configurations at the cost of a rather large increase in the error rate of the predictions.
Thus, although the combination of NegOW and OFAT is also on the Pareto frontier, the combination of OW and OFAT should be preferred, because it selects only a slightly larger number of configurations but substantially increases prediction accuracy.

We further analyzed performance-influence models for systems (e.g., HSMGP), for which we observe only a small difference in error rate when varying the binary sampling strategy.
Considering the learned influences of options and interactions, we see that interactions among binary options are negligible compared to the individual influences of the binary options.
Thus, even when using OW, it is possible to perform performance predictions with a small error rate.
Only when using RB(\_,OW), we have higher minimal error rates for these systems, in general, which can be explained by the property of random sampling of potentially not considering all binary configuration options in the sampling process.
Thus, all configurations with an unseen, influential option being enabled are predicted with a large error.

For numeric-sampling strategies, the main driver for prediction error is not the sampling strategy, but the sheer number of selected configurations.
That is, we see no systematic advantage of a certain strategy over another for the same number of configurations.
This partially contradicts further findings, which found the Plackett-Burman design as the most accurate sampling strategy~\cite{SGA+15}.
One reason for that behavior might be that a large part of the influences of the numeric configuration options can be described by simple polynomial functions (e.g., linear and quadratic functions), which can be revealed by all sampling strategies.
Furthermore, there seems to be no segmentations in the influences of the numeric options.
That is, one function can be used to describe the influence of an option over its whole value domain.
For example, linear functions can be used to describe the influence of the numeric configuration options of \textsc{TriMesh} over their whole value domain~\cite{GRS+16}.
In contrast, in some systems, there are options with different influences in disjoint parts of the value domain, as described by Ilyas et al.~\cite{ilyas_ea:europar:2017} or by Courtois and Woodside~\cite{DBLP:conf/wosp/CourtoisW00}. 
For these options, different functions have to be learned for different parts of their value domain.

For example, the numeric sampling strategy has only a small influence on prediction accuracy when considering the \textsc{gemm} subject system.
When analyzing its performance-influence models (models with an prediction error of less than 5\,\%), we see that none or, at most, only one of the numeric configuration options are part of the model.
Again, the small performance variation is a likely cause.

Having a closer look at the results for individual subject systems, we see that different sampling-strategy combinations are on the Pareto frontier.
This does not only hold for outlier systems, which we already discussed in Section~\ref{sec:result:bin:Out} and Section~\ref{sec:result:num:Out}, but for all systems.
We present the results for the individual subject systems at our supplementary Web site.
Overall, we find that no sampling strategy combination is on the Pareto frontier of all subject systems.
Instead, we see that 6 out of the 77 sampling-strategy combinations are one the Pareto frontier in half of our subject systems, while no sampling-strategy combination is optimal for more than three subject systems.
On our supplementary Web site, we provide a listing of all sampling-strategy combinations that are on at least one Pareto frontier, including the number of systems and the name of the systems for which the combination is on the frontier.
The reason for this finding is that the configuration options of the different systems have different performance influences on the systems.
Having a closer look at the sampling-strategy combinations, which are optimal for different systems, we do not observe common patterns across subject systems.

\Recommendation{
We recommend using a structured sampling strategy for binary configuration options, such as OW, T2, or T3 and we discourage using NegOW.
For numeric options, we recommend using sampling strategies that select a large number of configurations, depending on the available measurement budget.}

\subsubsection{Stability (\textsc{RQ2.2})}

Our experiments have shown that OW yields the most stable results regarding mean prediction error.
However, we also see that using sampling strategies from $\mathcal{B}_{\mathit{med}}$ or $\mathcal{B}_{\mathit{large}}$ can lead to smaller mean errors than from $\mathcal{B}_{\mathit{small}}$.
When analyzing some of the models that have been learned by MR, where using T2 leads to a smaller error rate compared to OW, we see that there are interactions among binary configuration options that are not part of the learning set produced by OW. 
Thus, parts of the larger distribution of the error rates achieved by the sampling strategies of $\mathcal{B}_{\mathit{med}}$ and $\mathcal{B}_{\mathit{large}}$ can be explained by OW's inability of identifying interactions among binary options.
However, especially for the sampling strategies of $\mathcal{B}_{\mathit{large}}$, we also see that, for some experiments, the error rates of $\mathcal{B}_{\mathit{large}}$ are larger compared to the error rates of $\mathcal{B}_{\mathit{small}}$.
Overall, we see this in 27\% of all of our experiments.
For the remaining 73\%, we see that sampling strategies of $\mathcal{B}_{\mathit{large}}$ lead to smaller error rates.
This is a surprising result as it contradicts the assumption that, with larger learning sets, we learn more accurate models. 
To learn more about this behavior, we analyzed one experiment where OW leads to a more accurate model compared to T3 when using MR in combination with CCI on \textsc{VP9}.
We found that the model of T3 considers a large number of interactions among binary configuration options and among binary and numeric configuration options compared to the other model based on OW.
In both models, we also see that the same options have been identified to influence the performance of the system.
Overfitting is a likely explanation of this observation. 
That is, the T3 models contain interactions that are merely measurement noise or wrongly classified higher-order interactions or influences of individual options that can not be derived correctly.

For numeric-sampling strategies, we do not see that selecting more configurations leads to more stable results for each experiment.
For example, in some experiments PBD(49,7) leads to more stable results compared to using PBD(125,5).

\Recommendation{To improve stability, we cannot recommend to improve the size of the learning set, in general.
For binary sampling, we see that a structured approach (i.e., OW, T2, and T3) provides more stable results than random sampling.
}

\subsection{Combinations of Machine-Learning Techniques and Sampling Strategies}

\subsubsection{Mean Error Rate (\textsc{RQ3.1})} \label{sec:discuss:r31}

The smallest error rates of our experiments are achieved when using RF as learning technique and RN(125) in combination with T3 as sampling strategies.
This is in line with our observations regarding the influence of the sampling strategies that larger learning sets lead to smaller error rates for a large number of experiments; and is also in line with the observations that mean error rates correlate with the number of selected configurations.
However, there is only a small difference in the error rates compared to using DOD(125) instead of RN(125).
When also considering the standard deviation of the prediction errors, DOD(125) should be favored, as this results in more stable predictions. 
Overall, this is in line with the results presented for RQ2.1.

\begin{figure}
  \centering
  \begin{subfigure}{1\textwidth}
  
   \begin{minipage}{0.65\textwidth}
          \centering
          \includegraphics[width=1\textwidth]{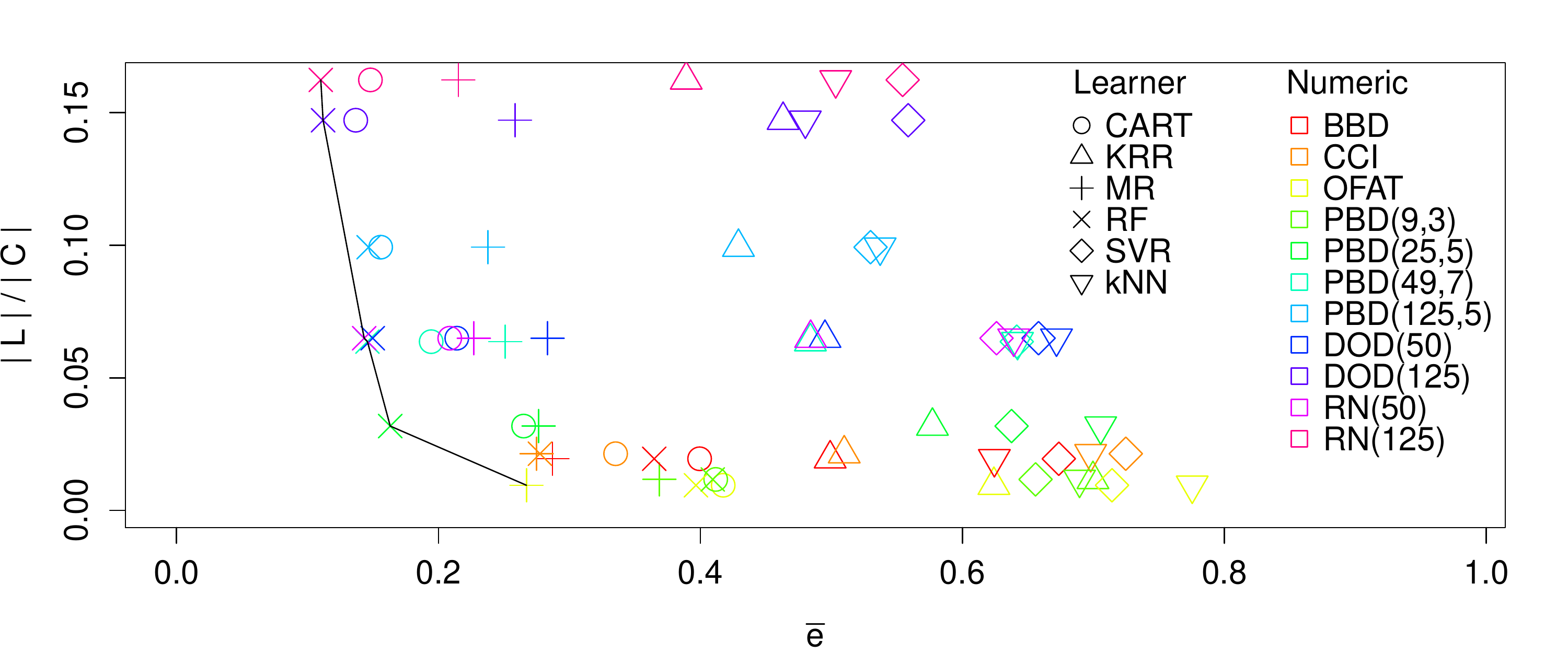}
          \caption{}
    \end{minipage}
    \begin{minipage}{0.3\textwidth}
      Sampling-strategy combinations on the Pareto frontier:\newline
      RF-RN(125), RF-DOD(125), RF-RN(50), RF-PBD(49,7), RF-PBD(25,5), and MR-OFAT;
      \vspace{1.3cm}
    \end{minipage}
  \end{subfigure}
  
  \begin{subfigure}{1\textwidth}
    \begin{minipage}{0.65\textwidth}
          \centering
          \includegraphics[width=1\textwidth]{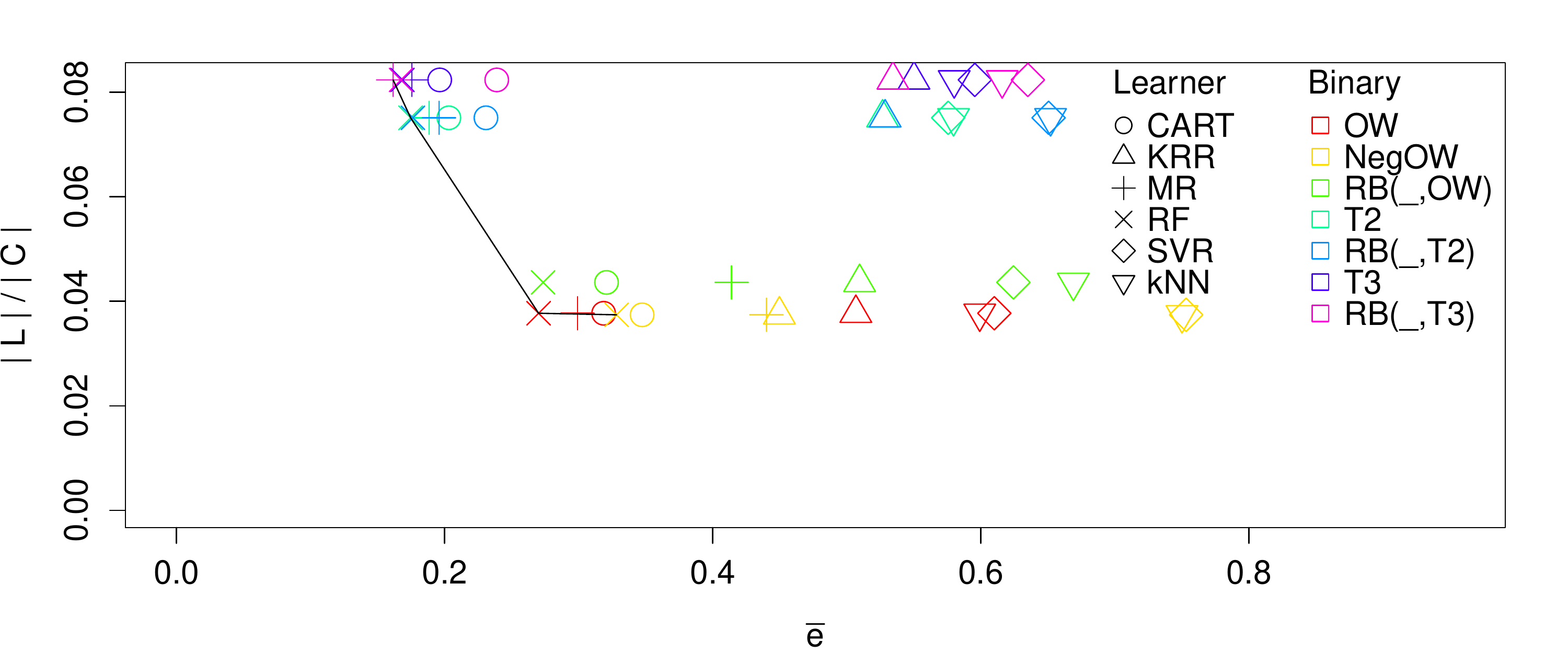}
          \caption{}
    \end{minipage}
    \begin{minipage}{0.3\textwidth}
      Sampling-strategy combinations on the Pareto frontier:\newline
      MR-RB(\_,T3), RF-T2, RF-OW, and RF-NegOW
      \vspace{2cm}
    \end{minipage}
\end{subfigure}
  \caption{Relative size of the learning set ($|L| / |C| $) and mean error rate ($\overline{e}$) (a) learning techniques and binary sampling strategies and (b) learning techniques and numeric sampling strategies aggregated over all subject systems. Next to the figures, we show the learning technique and sampling-strategy combinations of the Pareto frontiers ordered ascending by mean prediction error.
  }
  \label{fig:paretoFront_learner_XXX_average_rankingForBestConfig_FALSE}
\end{figure}

To provide a better overview of the relation between the error rate and the size of the learning set for the different learning techniques, in Figure~\ref{fig:paretoFront_learner_XXX_average_rankingForBestConfig_FALSE}a, we show the Pareto frontier of the learning techniques in combinations with the binary sampling strategies and, in Figure~\ref{fig:paretoFront_learner_XXX_average_rankingForBestConfig_FALSE}b, in combination with the numeric sampling strategies after aggregating over all subject systems.
We can see that, independent of the different sampling strategies, RF is able to achieve smaller error rates compared to the other learning techniques for almost all sampling sizes.
However, in Figure~\ref{fig:paretoFront_learner_XXX_average_rankingForBestConfig_FALSE}a, we see that MR outperforms RF for small sampling sizes.
Having a closer look at the results for the individual systems, we see that each of the considered learning techniques leads to the smallest mean error rate for, at least, one learning set size and subject system.
The results for the individual systems can be found on our supplementary Web site.

Overall, for almost all systems, it can be seen that, even with small learning sets (learning sets that were, for example, defined when using OW or NegOW), some learning techniques are able to predict all configurations with a small error rate.
For example, MR is able to predict all configurations of \textsc{TriMesh} with a mean error rate of 5.4\,\% when using OW in combination with BBD (which selects 150 configurations). 
Although the mean error rate of MR can be decreased to less than 4\,\% by, for example, applying T3 in combination with DOD(125), this requires 1850 configurations to be additionally measured. 
This drastic increase in measurement overhead leads to only a small decrease in the error rate.
This characteristic also holds for other systems and also other learning techniques.
So, although selecting a larger learning set often comes with benefits regarding prediction accuracy for a large number of systems, it is often a better tradeoff to accept a small loss in accuracy to avoid a substantial amount of measurement.

\Recommendation{
We can recommend to use RF in combination with structured sampling, because it leads to the most accurate predictions for the whole configuration space for a large number of experiments.
Besides that, we also see a tradeoff between error rate of the predictions and size of the learning set. 
Here, increasing the size of the learning set often does not drastically increase prediction accuracy.
As a consequence, we can recommend to select a small but representative learning set that leads to the best tradeoff among error rate and measurement overhead.
}

\subsubsection{Interactions among learning techniques and sampling strategies (\textsc{RQ3.2})}

To answer this research question, we search for outliers among the experiments that lead to predictions with a small error rate compared to applying the same machine-learning technique and other sampling strategies or vise versa.
Overall, we did not see specific interactions among the learning techniques and the sampling strategies that are not in line with the overall picture when considering all subject systems.

\Recommendation{
We did not observe strong interactions between machine-learning techniques and specific sampling strategies, so we cannot give a general recommendation.
}

\subsection{Summary}

In our experiments, we found that the error rate of the machine-learning techniques strongly depends on the subject system and the sampling strategy that is used.
Although we saw that some machine-learning techniques, such as CART and RF, are able to outperform other techniques such as SVR or KRR for a large number of learning sets and subject systems, the difference among the error rates of these two groups strongly depend on the subject system.
This leads to problems when comparing the efficiency of different machine-learning techniques when considering different subject systems, which is done in literature~\cite{SGS+15,Valov:2015:ECR:2791060.2791069,temple:hal-01467299}.
This problem becomes even more relevant when considering the influence of sampling strategies on the accuracy of different machine-learning techniques.

As a result, it is not possible to recommend one machine-learning technique outperforms over others independent to the learning set and the subject system being considered.


\section{Threats to Validity}

\subsection{Internal Validity}

In the literature, there is a large number of supervised machine-learning techniques that can be used to predict performance of all configurations of a configurable software system.
To provide meaningful recommendations, we select prominent techniques based on different strategies, such as trees-based strategies (CART or RF) or techniques that rely on a transformation of the input space (SVM or KRR).

To compare the machine-learning techniques in a fair manner, we perform a hyper-parameter optimization prior to predicting the performance of all configurations.
In this optimization, we consider parameters with a considerable influence of prediction accuracy.
Although we identified parameters leading to more accurate predictions compared to the default parameters, we cannot be certain whether we identified the optimal parameter values, because of the unbounded parameter spaces of the machine-learning techniques. 
To overcome this problem, we discretized parameters with real-valued parameter ranges and selected meaningful value domains that also consider the default values of the parameters. 

In our experiments, we use random sampling strategies for binary and the numeric configuration options to select a set of configurations of a specific size.
To quantify the influence of randomness, we applied the random selection several times using different seeds for the random number generator.
To this end, we consider the mean prediction accuracy when using different seeds.

For all subject systems, we measured all valid configurations to be able to consider the mean prediction accuracy for all configurations and not to be affected by the selection of a specific system.

\subsection{External Validity}

Configurable software systems are prevalent in many different domains, with varying sizes and complexity.
Not to be affected by domain-specific characteristics, we consider real-world software systems from different domains and of different size.
Overall, we consider configuration spaces ranging from roughly 2\,000 configurations to configuration spaces offering more than 200\,000 configurations.
Moreover, we consider run-time as well as compile-time configuration options in our experiments.
However, even by considering these different aspects, we can not simply transfer our results to other systems.
This can also be seen in our discussion about results on specific subject systems, where we saw that, for different systems, different machine-learning techniques or sampling strategies lead to more accurate predictions, which is one of our major findings here.
In the literature, there is a large number of machine-learning techniques and sampling strategies that can be used to solve the discusses problems. 
However, we select the most prominent learning techniques and sampling strategies representing the major approaches for learning a response function (performance model in our case).

\section{Related Work}

%

In previous work, it has been shown that machine-learning techniques can be used to accurately predict performance of software configurations and to identify the performance-optimal configuration with a high accuracy~\cite{Guo2013, SGA+15,Valov:2015:ECR:2791060.2791069}.
Most work proposes either using a specific learning technique or a suitbale sampling strategy.
To the best of our knowledge, a systematic comparison of the influence of different machine-learning techniques and sampling strategies on predicting performance for highly configurable software systems is missing.

In Section~\ref{sec:relWork:all}, we first discuss work that aims at predicting the performance of all configurations and, in Section~\ref{sec:relWork:opt}, we discuss work that aims at identifying the performance-optimal configuration.

\subsection{Performance Prediction of all Configurations}\label{sec:relWork:all}

Guo et al. use CART in combination with a random sampling strategy for performance prediction with a high accuracy~\cite{Guo2013}.
In a later work, they investigate progressive and projective sampling approaches to define the learning set~\cite{SGS+15}.
Here, they use sampling strategies based on feature frequency and the t-wise strategy to define the initial learning set. 
They found that projective sampling outperforms progressive sampling with respect to prediction accuracy.

In another line of research, Guo et al. propose DECART, data-efficient performance learning based on CART~\cite{GYS+17emse}.
Overall, this approach combines CART with resampling of configurations and a parameter tuning of the hyper-parameters of CART.
They show that DECART can be used the learn accurate performance models.
However, comparing this approach with other machine-learning approaches regarding the influence of different sampling strategies is hard, because one strength of this approach lies in actively increasing the size of the learning set in a structured way until certain termination criteria are reached.

Siegmund et al. use multivariable linear regression in combination with structured sampling strategies such as the Plackett Burmann Design to accurately predict performance of all configurations~\cite{SGA+15}.
They consider systems providing both binary and numeric options.
However, they consider only one machine-learning technique.
A similar machine-learning technique is used by Calotoiu et al.~\cite{calotoiu_ea:2016}.
They aim at learning the influence of different options on the execution time of parameters such as the number of CPU cores on the performance of specific parts of software systems from the high performance computing domain.
To learn these influences, they use a full factorial design by considering 5 different values of each of the options; they do not consider other sampling strategies.
In contrast to our work, they consider only a small number of configuration options in their models.

Nair et al. apply spectral learning to reduce the size of the learning set, while predicting all configurations with a high accuracy~\cite{DBLP:journals/corr/NairMSA17}.
To this end, they use spectral sampling to define the learning set.
They compare their approach with CART, multivariable linear regression, and progressive sampling used by Sarkar et al.~\cite{SGS+15}.
They found that their approach suffices with a smaller learning set compared to the other approaches while maintaining accuracy. 
However, they have a strong coupling between their machine-learning technique and the sampling strategy, which makes it hard to use their sampling strategy with other learning techniques.

Valov et al.~\cite{Valov:2015:ECR:2791060.2791069} compared different regression techniques, including CART, RF, and SVR.
In their experiments, they perform a parameter optimization to identify the best, average, and worst parameters with respect to accuracy of the predictions.
These parameters are then used to predict performance of all configurations.
Moreover, they also consider the size of the learning set in their experiments.
In contrast to our work, they consider only binary configuration options and apply only random sampling to define the learning set. 
Although the bagging technique, which they consider in their comparison, produces the best results regarding prediction accuracy, we omit this technique in our experiments, because we consider machine-learning techniques that use different strategies internally (e.g., tree-based techniques and techniques that use an input-value transformations, such as SVR or KRR).

For some systems, it has been noticed that the influence of configuration options can not described using continuous functions.
One reason for this are internal changes of the processing of data depending on the value of configuration options~\cite{DBLP:conf/wosp/CourtoisW00}. 
As a consequence, it is not possible to describe the influence of the option with one function over the complete value domain of the option. 
To address this problem, one can learn spline functions, where different splines describe the influence of different parts of the value domain~\cite{DBLP:conf/wosp/CourtoisW00, DBLP:conf/wosp/NoorshamsBKR13}.

In another line of research, Faber and Happe~\cite{DBLP:conf/wosp/FaberH12} use genetic programming to derive performance models in terms of performance curves.
These performance curves describe the influence of configuration options and their interactions on the performance of the system and thus, are semantically equivalent to the performance-influence models learned by the MR technique.
However, instead of iteratively traversing the configuration space and using regression, the authors use crossover to combine two performance curves and mutation to adjust the influence of the configuration options.
They found out that the accuracy of such an approach strongly depends on it's parameters and the problem to consider.
To increase accuracy of the curves, they integrated domain knowledge in the model derivation process. 
However, their domain knowledge does not have an influence on the results.

\subsection{Identifying the Performance-Optimal Configuration}\label{sec:relWork:opt}

Temple et al. use CART to identify performance-optimal configurations~\cite{temple:hal-01467299}.
With a classification tree, they separate configurations with a good performance from configurations with bad performance.
To define the learning set, they use random sampling.
In their evaluation, they consider 16 non-functional properties of different configurable systems.
In contrast to our work, they use the classification tree to separate the configurations based on their performance and do not consider structured sampling strategies.

In another line of research, Nair et al.~\cite{DBLP:journals/corr/abs-1801-02175} use a sequential model-based method to identify to performance optimal configurations.
For sampling, they start with a random sampling of configurations and refine the set by additional configurations based on different heuristics.

A completely different approach to identify near-optimal configurations is used by Oh et al.~\cite{DBLP:conf/sigsoft/OhBMS17}. 
Instead of using a machine-learning technique, which aims at learning a model, they iteratively shrink the configuration space towards the performance-optimal variant using information given in a random sample in each iteration step.
They obtain this random set of configurations by applying counting binary decision diagrams, which are binary decision diagrams that also store the number of possible configurations at each node of the diagram, on the configuration space of a system.

Nair et al. shows that it is not necessary to learn an accurate performance model to identify optimal, or at least, nearly optimal configurations~\cite{NMS+17}.
They further show that a rank-based approach can be used to drastically reduce the number of configurations required to measure as well as the time to compute the model compared to considering the average error rate of the configurations.

Zhang et al. proposes an approach based on Fourier learning~\cite{DBLP:conf/kbse/ZhangGBC15}.
Using this approach, it is possible to provide theoretical grantees of accuracy.
However, this approach only works when the performance function is in Fourier space.

\section{Conclusion}

Predicting the performance of the configurations of configurable software system is a non-trivial task.
To overcome this problem, supervised machine-learning techniques can be used, which rely on a learning set of measured configurations.
In the literature, there is a large number of such learning techniques, which differ in the concepts and mechanisms they use to learn a model to predict the performance of the configurations.
There is also a large number of sampling strategies that can be used to define a learning set.
This makes it hard to identify the combination of machine-learning technique and sampling strategy that leads to the most accurate predictions, either when predicting the performance-optimal configuration or the performance of all configurations.

To shed light into this problem, we conducted an empirical study to identify the influence of selecting individual machine-learning techniques and sampling strategies on the accuracy of performance predictions.
Specifically, we compare 6 machine-learning techniques and 18 sampling strategies (for binary and numeric configuration options) regarding their influence on the mean error rate when predicting all configurations.
In the study, we use \changeReview{6} real world configurable subject systems from different domains that offer runtime as well as compile-time configuration options.

In our experiments, we found that the selection of machine-learning technique and sampling strategy has a strong influence on the accuracy of performance predictions.
For the machine-learning techniques, we saw that three of the six techniques (CART, MR, and RF) led in many cases to substantially more accurate predictions (compared to kNN, KRR, and SVR).
However, when considering individual software systems, we also see that the difference in the prediction accuracy depend not only on the sampling strategy, but also on the specifics of the considered software system (for example, for DUNE MGS, we see only small differences in the prediction accuracy when using different machine-learning techniques).
For binary sampling strategies, we saw that applying structured sampling strategies leads to more accurate predictions compared to using a random learning set, especially when selecting only a small number of configurations.
We also found that this difference decreases with an increasing number of selected configurations.
By contrast, for the selection of the numeric options, we found that the accuracy depends on the number of selected configurations and only at a small extent on the learning set.
Overall, we can say that for predicting the performance of all configurations, the selection of the machine-learning technique is more relevant.



\section*{Acknowledgments}

This work is supported by the German Research Foundation (DFG), as part of the Priority Programme 1648 ``Software for Exascale Computing'', under the contract AP 206/7.
Apel's work is supported by the DFG under the contracts AP 206/4, AP 206/6, and AP 206/7. 
Siegmund's work is supported by the DFG under the contracts SI 2171/2 and SI 2171/3-1.



\bibliographystyle{IEEEtran}
\bibliography{ms}

\begin{thebibliography}{10}
\providecommand{\url}[1]{#1}
\csname url@samestyle\endcsname
\providecommand{\newblock}{\relax}
\providecommand{\bibinfo}[2]{#2}
\providecommand{\BIBentrySTDinterwordspacing}{\spaceskip=0pt\relax}
\providecommand{\BIBentryALTinterwordstretchfactor}{4}
\providecommand{\BIBentryALTinterwordspacing}{\spaceskip=\fontdimen2\font plus
\BIBentryALTinterwordstretchfactor\fontdimen3\font minus
  \fontdimen4\font\relax}
\providecommand{\BIBforeignlanguage}[2]{{%
\expandafter\ifx\csname l@#1\endcsname\relax
\typeout{** WARNING: IEEEtran.bst: No hyphenation pattern has been}%
\typeout{** loaded for the language `#1'. Using the pattern for}%
\typeout{** the default language instead.}%
\else
\language=\csname l@#1\endcsname
\fi
#2}}
\providecommand{\BIBdecl}{\relax}
\BIBdecl

\bibitem{SGA+15}
N.~Siegmund, A.~Grebhahn, S.~Apel, and C.~K{\"a}stner, ``{Performance-Influence
  Models for Highly Configurable Systems},'' in \emph{Proceedings of the Joint
  Meeting of the European Software Engineering Conference and the ACM SIGSOFT
  International Symposium on the Foundations of Software Engineering
  (ESEC/FSE)}.\hskip 1em plus 0.5em minus 0.4em\relax ACM, 2015, pp. 284--294.

\bibitem{Xu:2015:HYG:2786805.2786852}
T.~Xu, L.~Jin, X.~Fan, Y.~Zhou, S.~Pasupathy, and R.~Talwadker, ``{Hey, You
  Have Given Me Too Many Knobs!: Understanding and Dealing with Over-designed
  Configuration in System Software},'' in \emph{Proceedings of the Joint
  Meeting of the European Software Engineering Conference and the ACM SIGSOFT
  International Symposium on the Foundations of Software Engineering
  (ESEC/FSE)}.\hskip 1em plus 0.5em minus 0.4em\relax ACM, 2015, pp. 307--319.

\bibitem{DBLP:conf/ics/GahvariBSYJG11}
H.~Gahvari, A.~H. Baker, M.~Schulz, U.~M. Yang, K.~E. Jordan, and W.~Gropp,
  ``{Modeling the Performance of an Algebraic Multigrid Cycle on {HPC}
  Platforms},'' in \emph{Proceedings of the International Conference on
  Supercomputing (ICS)}, 2011, pp. 172--181.

\bibitem{books/daglib/0076234}
R.~Jain, \emph{{The Art of Computer Systems Performance Analysis - Techniques
  for Experimental Design, Measurement, Simulation, and Modeling.}}, ser. Wiley
  professional computing.\hskip 1em plus 0.5em minus 0.4em\relax Wiley, 1991.

\bibitem{Guo2013}
J.~Guo, K.~Czarnecki, S.~Apel, N.~Siegmund, and A.~Wasowski,
  ``{Variability-Aware Performance Prediction: A Statistical Learning
  Approach},'' in \emph{Proceedings of the International Conference on
  Automated Software Engineering (ASE)}.\hskip 1em plus 0.5em minus 0.4em\relax
  ACM, 2013, pp. 301--311.

\bibitem{cart84}
L.~Breiman, J.~Friedman, R.~Olshen, and C.~Stone, \emph{{Classification and
  Regression Trees}}.\hskip 1em plus 0.5em minus 0.4em\relax Monterey, CA:
  Wadsworth and Brooks, 1984.

\bibitem{basak2007support}
D.~Basak, S.~Pal, and D.~C. Patranabis, ``{Support Vector Regression},''
  \emph{Neural Information Processing-Letters and Reviews}, vol.~11, no.~10,
  pp. 203--224, 2007.

\bibitem{anderson2003introduction}
T.~Anderson, \emph{{An Introduction to Multivariate Statistical Analysis}},
  ser. Wiley Series in Probability and Statistics.\hskip 1em plus 0.5em minus
  0.4em\relax Wiley, 2003.

\bibitem{KSK+18}
S.~Kolesnikov, N.~Siegmund, C.~K{\"a}stner, A.~Grebhahn, and S.~Apel,
  ``{Tradeoffs in Modeling Performance of Highly-Configurable Software
  Systems},'' \emph{Software and Systems Modeling (SoSyM)}, pp. 1--19, 2018,
  online first.

\bibitem{Valov:2015:ECR:2791060.2791069}
P.~Valov, J.~Guo, and K.~Czarnecki, ``{Empirical Comparison of Regression
  Methods for Variability-aware Performance Prediction},'' in \emph{Proceedings
  of the Software Product Line Conference (SPLC)}.\hskip 1em plus 0.5em minus
  0.4em\relax ACM, 2015, pp. 186--190.

\bibitem{Garcia-Gutierrez2014}
J.~Garc{\'{\i}}a{-}Guti{\'{e}}rrez, F.~Mart{\'{\i}}nez{-}{\'{A}}lvarez, A.~T.
  Lora, and J.~C. Riquelme, ``{A Comparative Study of Machine Learning
  Regression Methods on LiDAR Data: {A} Case Study},'' in \emph{International
  Joint Conference SOCO'13-CISIS'13-ICEUTE'13 - Salamanca}, 2013, pp. 249--258.

\bibitem{KaGrSi+19}
C.~Kaltenecker, A.~Grebhahn, N.~Siegmund, J.~Guo, and S.~Apel,
  ``{Distance-Based Sampling of Software Configuration Spaces},'' in
  \emph{Proceedings of the International Conference on Software Engineering
  (ICSE)}.\hskip 1em plus 0.5em minus 0.4em\relax IEEE Computer Society, 2019.

\bibitem{SGS+15}
A.~Sarkar, J.~Guo, N.~Siegmund, S.~Apel, and K.~Czarnecki, ``{Cost-Efficient
  Sampling for Performance Prediction of Configurable Systems},'' in
  \emph{Proceedings of the International Conference on Automated Software
  Engineering (ASE)}.\hskip 1em plus 0.5em minus 0.4em\relax IEEE, 2015, pp.
  342--352.

\bibitem{PlackettBurman1946}
R.~L. Plackett and J.~P. Burman, ``\BIBforeignlanguage{English}{{The Design of
  Optimum Multifactorial Experiments}},''
  \emph{\BIBforeignlanguage{English}{Biometrika}}, vol.~33, no.~4, pp.
  305--325, 1946.

\bibitem{DBLP:journals/corr/abs-1801-02175}
V.~Nair, Z.~Yu, T.~Menzies, N.~Siegmund, and S.~Apel, ``{ Finding Faster
  Configurations using FLASH},'' \emph{{IEEE Transactions on Software
  Engineering (TSE)}}, 2018, online first.

\bibitem{DBLP:journals/corr/NairMSA17}
V.~Nair, T.~Menzies, N.~Siegmund, and S.~Apel, ``{Faster Discovery of Faster
  System Configurations with Spectral Learning},'' \emph{Automated Software
  Engineering}, vol.~25, no.~2, pp. 247--277, 2018.

\bibitem{temple:hal-01467299}
P.~A. Temple, M.~Acher, J.-M.~A. J{\'e}z{\'e}quel, L.~A. Noel-Baron, and J.~A.
  Galindo, ``{Learning-Based Performance Specialization of Configurable
  Systems},'' {IRISA, Inria Rennes; University of Rennes 1}, Research Report,
  2017.

\bibitem{myers2009response}
R.~Myers, D.~Montgomery, and C.~Anderson-Cook, \emph{{Response Surface
  Methodology: Process and Product Optimization Using Designed Experiments}},
  ser. Wiley Series in Probability and Statistics.\hskip 1em plus 0.5em minus
  0.4em\relax Wiley, 2009.

\bibitem{SSA+17}
N.~Siegmund, S.~Sobernig, and S.~Apel, ``{Attributed Variability Models:
  Outside the Comfort Zone},'' in \emph{Proceedings of the Joint Meeting of the
  European Software Engineering Conference and the ACM SIGSOFT International
  Symposium on the Foundations of Software Engineering (ESEC/FSE)}.\hskip 1em
  plus 0.5em minus 0.4em\relax ACM, 2017, pp. 268--278.

\bibitem{DBLP:conf/sigsoft/FuM17}
W.~Fu and T.~Menzies, ``{Easy over Hard: a Case Study on Deep Learning},'' in
  \emph{Proceedings of the Joint Meeting of the European Software Engineering
  Conference and the ACM SIGSOFT International Symposium on the Foundations of
  Software Engineering (ESEC/FSE)}, 2017, pp. 49--60.

\bibitem{loh2011classification}
W.-Y. Loh, ``{Classification and Regression Trees},'' \emph{Wiley
  Interdisciplinary Reviews: Data Mining and Knowledge Discovery}, vol.~1,
  no.~1, pp. 14--23, 2011.

\bibitem{Breiman2001}
L.~Breiman, ``{Random Forests},'' \emph{Machine Learning}, vol.~45, no.~1, pp.
  5--32, 2001.

\bibitem{vapnik1995nature}
V.~Vapnik, ``{The Nature of Statistical Learning Theory},'' 1995.

\bibitem{smola2004tutorial}
A.~J. Smola and B.~Sch{\"o}lkopf, ``{A Tutorial on Support Vector
  Regression},'' \emph{{Statistics and Computing}}, vol.~14, no.~3, pp.
  199--222, 2004.

\bibitem{Vovk2013}
V.~Vovk, \emph{{Kernel Ridge Regression}}.\hskip 1em plus 0.5em minus
  0.4em\relax Springer, 2013, pp. 105--116.

\bibitem{Cristianini:1999:ISV:345662}
N.~Cristianini and J.~Shawe-Taylor, \emph{{An Introduction to Support Vector
  Machines: And Other Kernel-based Learning Methods}}.\hskip 1em plus 0.5em
  minus 0.4em\relax Cambridge University Press, 2000.

\bibitem{Murphy:2012:MLP:2380985}
K.~P. Murphy, \emph{{Machine Learning: A Probabilistic Perspective}}.\hskip 1em
  plus 0.5em minus 0.4em\relax The MIT Press, 2012.

\bibitem{Mitchell:1997:ML:541177}
T.~M. Mitchell, \emph{{Machine Learning}}, 1st~ed.\hskip 1em plus 0.5em minus
  0.4em\relax McGraw-Hill, Inc., 1997.

\bibitem{doi:10.1080/00031305.1992.10475879}
N.~S. Altman, ``{An Introduction to Kernel and Nearest-Neighbor Nonparametric
  Regression},'' \emph{The American Statistician}, vol.~46, no.~3, pp.
  175--185, 1992.

\bibitem{Medeiros:2016:CSA:2884781.2884793}
F.~Medeiros, C.~K\"{a}stner, M.~Ribeiro, R.~Gheyi, and S.~Apel, ``{A Comparison
  of 10 Sampling Algorithms for Configurable Systems},'' in \emph{Proceedings
  of the International Conference on Software Engineering (ICSE)}.\hskip 1em
  plus 0.5em minus 0.4em\relax ACM, 2016, pp. 643--654.

\bibitem{ARW+13}
S.~Apel, A.~von Rhein, P.~Wendler, A.~Gr{\"o}{\ss}linger, and D.~Beyer,
  ``{Strategies for Product-Line Verification: Case Studies and Experiments},''
  in \emph{Proceedings of the International Conference on Software Engineering
  (ICSE)}.\hskip 1em plus 0.5em minus 0.4em\relax IEEE, 2013, pp. 482--491.

\bibitem{Perrouin:2010:AST:1828417.1828490}
G.~Perrouin, S.~Sen, J.~Klein, B.~Baudry, and Y.~l. Traon, ``{Automated and
  Scalable T-wise Test Case Generation Strategies for Software Product
  Lines},'' in \emph{Proceedings of the International Conference on Software
  Testing, Verification and Validation (ICST)}.\hskip 1em plus 0.5em minus
  0.4em\relax IEEE, 2010, pp. 459--468.

\bibitem{Nie:2011:SCT:1883612.1883618}
C.~Nie and H.~Leung, ``{A Survey of Combinatorial Testing},'' \emph{ACM
  Computing Surveys}, vol.~43, no.~2, pp. 11:1--11:29, 2011.

\bibitem{Chakraborty:2014:DSW:2892753.2892792}
S.~Chakraborty, D.~J. Fremont, K.~S. Meel, S.~A. Seshia, and M.~Y. Vardi,
  ``{Distribution-aware Sampling and Weighted Model Counting for SAT},'' in
  \emph{Proceedings of the AAAI Conference on Artificial Intelligence
  (AAAI)}.\hskip 1em plus 0.5em minus 0.4em\relax AAAI Press, 2014, pp.
  1722--1730.

\bibitem{10.2307/2685731}
V.~Czitrom, ``{One-Factor-at-a-Time versus Designed Experiments},'' \emph{The
  American Statistician}, vol.~53, no.~2, pp. 126--131, 1999.

\bibitem{morris1991factorial}
M.~D. Morris, ``{Factorial Sampling Plans for Preliminary Computational
  Experiments},'' \emph{Technometrics}, vol.~33, no.~2, pp. 161--174, 1991.

\bibitem{campolongo2007effective}
F.~Campolongo, J.~Cariboni, and A.~Saltelli, ``{An Effective Screening Design
  for Sensitivity Analysis of Large Models},'' \emph{{Environmental Modelling
  \& Software}}, vol.~22, no.~10, pp. 1509--1518, 2007.

\bibitem{BoxBehnkenDesign}
G.~E.~P. Box and D.~W. Behnken, ``{Some New Three Level Designs for the Study
  of Quantitative Variables},'' \emph{Technometrics}, vol.~2, no.~4, pp.
  455--475, 1960.

\bibitem{Montgomery:2006:DAE:1206386}
D.~C. Montgomery, \emph{{Design and Analysis of Experiments}}.\hskip 1em plus
  0.5em minus 0.4em\relax John Wiley \& Sons, 2001.

\bibitem{ferreira2007box}
S.~C. Ferreira, R.~Bruns, H.~Ferreira, G.~Matos, J.~David, G.~Brandao, E.~P.
  da~Silva, L.~Portugal, P.~Dos~Reis, A.~Souza \emph{et~al.}, ``{Box-Behnken
  Design: An Alternative for the Optimization of Analytical Methods},''
  \emph{Analytica Chimica Acta}, vol. 597, no.~2, pp. 179--186, 2007.

\bibitem{Nguyen2008294}
N.-K. Nguyen and J.~J. Borkowski, ``{New 3-level Response Surface Designs
  Constructed from Incomplete Block Designs},'' \emph{Journal of Statistical
  Planning and Inference}, vol. 138, no.~1, pp. 294 -- 305, 2008.

\bibitem{ASLAN200790}
N.~Aslan and Y.~Cebeci, ``{Application of Box-Behnken Design and Response
  Surface Methodology for Modeling of some Turkish Coals},'' \emph{Fuel},
  vol.~86, no.~1, pp. 90 -- 97, 2007.

\bibitem{Annadurai1998}
G.~Annadurai and R.~Y. Sheeja, ``{Use of Box-Behnken Design of Experiments for
  the Adsorption of Verofix red using Biopolymer},'' \emph{Bioprocess
  Engineering}, vol.~18, no.~6, pp. 463--466, 1998.

\bibitem{BoxWilson1951}
G.~E.~P. Box and K.~B. Wilson, ``\BIBforeignlanguage{English}{{On the
  Experimental Attainment of Optimum Conditions}},''
  \emph{\BIBforeignlanguage{English}{Journal of the Royal Statistical Society.
  Series B (Methodological)}}, vol.~13, no.~1, pp. 1--45, 1951.

\bibitem{KhuriMukho10}
A.~I. Khuri and S.~Mukhopadhyay, ``{Response Surface Methodology},''
  \emph{Wiley Interdisciplinary Reviews: Computational Statistics}, vol.~2,
  no.~2, pp. 128--149, 2010.

\bibitem{yeten2005comparison}
B.~Yeten, A.~Castellini, B.~Guyaguler, and W.~H. Chen, ``{A Comparison Study on
  Experimental Design and Response Surface Methodologies},'' in \emph{SPE
  Reservoir Simulation Symposium}.\hskip 1em plus 0.5em minus 0.4em\relax
  Society of Petroleum Engineers, 2005.

\bibitem{garg2013selection}
A.~Garg and K.~Tai, ``{Selection of a Robust Experimental Design for the
  Effective Modeling of Nonlinear Systems using Genetic Programming},'' in
  \emph{{IEEE} Symposium on Computational Intelligence and Data Mining
  (CIDM)}.\hskip 1em plus 0.5em minus 0.4em\relax IEEE, 2013, pp. 287--292.

\bibitem{Jacques1999}
P.~Jacques, C.~Hbid, J.~Destain, H.~Razafindralambo, M.~Paquot, E.~De~Pauw, and
  P.~Thonart, ``{Optimization of Biosurfactant Lipopeptide Production from
  Bacillus subtilis S499 by Plackett-Burman Design},'' \emph{Applied
  Biochemistry and Biotechnology}, vol.~77, no.~1, pp. 223--233, 1999.

\bibitem{Krishnan1998}
S.~Krishnan, S.~G. Prapulla, D.~Rajalakshmi, M.~C. Misra, and N.~G. Karanth,
  ``{Screening and Selection of Media Components for Lactic Acid Production
  using Plackett--Burman Design},'' \emph{Bioprocess Engineering}, vol.~19,
  no.~1, pp. 61--65, 1998.

\bibitem{Smith1918}
K.~Smith, ``\BIBforeignlanguage{English}{{On the Standard Deviations of
  Adjusted and Interpolated Values of an Observed Polynomial Function and its
  Constants and the Guidance they give Towards a Proper Choice of the
  Distribution of Observations}},''
  \emph{\BIBforeignlanguage{English}{Biometrika}}, vol.~12, no. 1/2, pp. 1--85,
  1918.

\bibitem{fedorov1972theory}
V.~Fedorov, \emph{{Theory of Optimal Experiments}}, ser. Probability and
  Mathematical Statistics.\hskip 1em plus 0.5em minus 0.4em\relax Academic
  Press, 1972.

\bibitem{mitchellExchangeAlgorithm}
T.~J. Mitchell, ``{An Algorithm for the Construction of "D-Optimal"
  Experimental Designs},'' \emph{Technometrics}, vol.~16, no.~2, pp. 203--210,
  1974.

\bibitem{kexchange}
M.~E. Johnson and C.~J. Nachtsheim, ``{Some Guidelines for Constructing Exact
  D-Optimal Designs on Convex Design Spaces},'' \emph{Technometrics}, vol.~25,
  no.~3, pp. 271--277, 1983.

\bibitem{FANG2011717}
S.-E. Fang and R.~Perera, ``{Damage Identification by Response Surface based
  Model updating using D-Optimal Design},'' \emph{Mechanical Systems and Signal
  Processing}, vol.~25, no.~2, pp. 717 -- 733, 2011.

\bibitem{Ganapathi:2009:CML:1855591.1855592}
A.~Ganapathi, K.~Datta, A.~Fox, and D.~Patterson, ``{A Case for Machine
  Learning to Optimize Multicore Performance},'' in \emph{Proceedings of the
  USENIX Conference on Hot Topics in Parallelism}.\hskip 1em plus 0.5em minus
  0.4em\relax USENIX Association, 2009, pp. 1--6.

\bibitem{bergstra2012machine}
J.~Bergstra, N.~Pinto, and D.~Cox, ``{Machine Learning for Predictive
  Auto-tuning with Boosted Regression Trees},'' in \emph{Innovative Parallel
  Computing (InPar), 2012}.\hskip 1em plus 0.5em minus 0.4em\relax IEEE, 2012,
  pp. 1--9.

\bibitem{saltelli2008global}
A.~Saltelli, M.~Ratto, T.~Andres, F.~Campolongo, J.~Cariboni, D.~Gatelli,
  M.~Saisana, and S.~Tarantola, \emph{{Global Sensitivity Analysis: The
  Primer}}.\hskip 1em plus 0.5em minus 0.4em\relax John Wiley \& Sons, 2008.

\bibitem{DBLP:journals/infsof/FuMS16}
W.~Fu, T.~Menzies, and X.~Shen, ``{Tuning for Software Analytics: Is it Really
  Necessary?}'' \emph{Information {\&} Software Technology}, vol.~76, pp.
  135--146, 2016.

\bibitem{Bergstra:2012:RSH:2188385.2188395}
J.~Bergstra and Y.~Bengio, ``{Random Search for Hyper-parameter
  Optimization},'' \emph{Journal of Machine Learning Research}, vol.~13, pp.
  281--305, 2012.

\bibitem{Cliff1996}
N.~Cliff, \emph{{Ordinal Methods for Behavioral Data Analysis}}.\hskip 1em plus
  0.5em minus 0.4em\relax Erlbaum, 1996.

\bibitem{DBLP:conf/sigsoft/OhBMS17}
J.~Oh, D.~S. Batory, M.~Myers, and N.~Siegmund, ``{Finding Near-Optimal
  Configurations in Product Lines by Random Sampling},'' in \emph{Proceedings
  of the Joint Meeting of the European Software Engineering Conference and the
  ACM SIGSOFT International Symposium on the Foundations of Software
  Engineering (ESEC/FSE)}, 2017, pp. 61--71.

\bibitem{GRS+16}
A.~Grebhahn, C.~Rodrigo, N.~Siegmund, F.~J. Gaspar, and S.~Apel,
  ``{Performance-Influence Models of Multigrid Methods: A Case Study on
  Triangular Meshes},'' \emph{Concurrency and Computation: Practice and
  Experience}, vol.~29, no.~17, pp. 4057:1--4057:13, 2017.

\bibitem{ilyas_ea:europar:2017}
K.~Ilyas, A.~Calotoiu, and F.~Wolf, ``{Off-Road Performance Modeling -- How to
  Deal with Segmented Data},'' in \emph{Proceedings of the Euro-Par Conference
  (Euro-Par)}, ser. Lecture Notes in Computer Science.\hskip 1em plus 0.5em
  minus 0.4em\relax Springer, 2017, pp. 36--48.

\bibitem{DBLP:conf/wosp/CourtoisW00}
M.~Courtois and C.~M. Woodside, ``{Using Regression Splines for Software
  Performance Analysis},'' in \emph{Workshop on Software and Performance},
  2000, pp. 105--114.

\bibitem{GYS+17emse}
J.~Guo, D.~Yang, N.~Siegmund, S.~Apel, A.~Sarkar, P.~Valov, K.~Czarnecki,
  A.~Wasowski, and H.~Yu, ``{Data-Efficient Performance Learning for
  Configurable Systems},'' \emph{Empirical Software Engineering}, vol.~23,
  no.~3, pp. 1826--1867, 2018.

\bibitem{calotoiu_ea:2016}
A.~Calotoiu, D.~Beckingsale, C.~W. Earl, T.~Hoefler, I.~Karlin, M.~Schulz, and
  F.~Wolf, ``{Fast Multi-Parameter Performance Modeling},'' in
  \emph{Proceedings of the IEEE International Conference on Cluster Computing
  (CLUSTER)}.\hskip 1em plus 0.5em minus 0.4em\relax IEEE, 2016, pp. 172--181.

\bibitem{DBLP:conf/wosp/NoorshamsBKR13}
Q.~Noorshams, D.~Bruhn, S.~Kounev, and R.~H. Reussner, ``{Predictive
  Performance Modeling of Virtualized Storage Systems using Optimized
  Statistical Regression Techniques},'' in \emph{International Conference on
  Performance Engineering (ICPE)}, 2013, pp. 283--294.

\bibitem{DBLP:conf/wosp/FaberH12}
M.~Faber and J.~Happe, ``{Systematic Adoption of Genetic Programming for
  Deriving Software Performance Curves},'' in \emph{International Conference on
  Performance Engineering (ICPE)}, 2012, pp. 33--44.

\bibitem{NMS+17}
V.~Nair, T.~Menzies, N.~Siegmund, and S.~Apel, ``{Using Bad Learners to find
  Good Configurations},'' in \emph{Proceedings of the Joint Meeting of the
  European Software Engineering Conference and the ACM SIGSOFT International
  Symposium on the Foundations of Software Engineering (ESEC/FSE)}.\hskip 1em
  plus 0.5em minus 0.4em\relax ACM Press, 2017, pp. 257--267.

\bibitem{DBLP:conf/kbse/ZhangGBC15}
Y.~Zhang, J.~Guo, E.~Blais, and K.~Czarnecki, ``{Performance Prediction of
  Configurable Software Systems by Fourier Learning {(T)}},'' in
  \emph{Proceedings of the International Conference on Automated Software
  Engineering (ASE)}, 2015, pp. 365--373.

\end{thebibliography}

%

\end{document}